\documentclass[a4paper,11pt,pdf]{article}

\usepackage{amsfonts}
\usepackage{amsmath}
\usepackage{amsthm}
\usepackage[auth-sc,affil-sl]{authblk}
\usepackage{bm}
\usepackage{multirow}
\usepackage{verbatim}
\usepackage{enumitem}
\usepackage{hyperref}

\renewcommand
\baselinestretch{1.0} 
\usepackage{url}
\usepackage{natbib}
\bibpunct{(}{)}{,}{a}{}{,}
\usepackage{graphicx}
\graphicspath{{figures/}} 
\usepackage{lscape}
\usepackage[svgnames]{xcolor}
\usepackage{sansmath}
\usepackage{subfigure}
\usepackage{rotating}

\DeclareMathAlphabet{\mathpzc}{OT1}{pzc}{m}{it}

\def\transpose{{\sf{T}}}

\def\V{\mathrm{Var}}

\newcommand\independent{\protect\mathpalette{\protect\independenT}{\perp}}
\def\independenT#1#2{\mathrel{\rlap{$#1#2$}\mkern2mu{#1#2}}}

\def\Expect{\mathbb{{E}}}
\def\P{\text{Pr}}
\def\P{\mathrm{Pr}}
\def\Cov{\text{Cov}}

\def\plim{\mathrm{plim}}

\def\b0{\boldsymbol{0}}

\newtheorem{Assum}{Assumption}
\newtheorem{Def}{Definition}
\newtheorem{Lemma}{Lemma}
\newtheorem{Proof}{Proof}
\newtheorem{Thm}{Theorem}
\newtheorem{Corol}{Corollary}

\xdefinecolor{myblue}{rgb}{0,0.2,0.5}
\xdefinecolor{impblue}{rgb}{0.2,0.3,0.5}
\xdefinecolor{myred}{rgb}{0.7,0.1,0.2}
\xdefinecolor{mygreen}{rgb}{0,0.30,0}

\newcommand \address[1]{\gdef \@address{#1}}
\makeatletter
\long\def\@footnotetext#1{\insert\footins{\def\baselinestretch{1.2}\footnotesize
\interlinepenalty\interfootnotelinepenalty
\splittopskip\footnotesep \splitmaxdepth \dp\strutbox
\floatingpenalty \@MM \hsize\columnwidth \@parboxrestore
\edef\@currentlabel{\csname
p@footnote\endcsname\@thefnmark}\@makefntext
{\rule{\z@}{\footnotesep}\ignorespaces #1\strut}}}

\long\def\symbolfootnote[#1]#2{\begingroup%
\def\thefootnote{\fnsymbol{footnote}}\footnote[#1]{#2}\endgroup}

\def\maketitle{%
  \null
  \thispagestyle{empty}%
  \begin{center}\leavevmode
    \normalfont
    {\LARGE \bf \@title\par}%
    {\normalsize \@author\par}%
    \vskip 0.05 cm
    \vskip 0.05cm
    {\normalsize \@date\par}%
  \end{center}%
}

\makeatletter
\newcommand{\institute}[1]{\newcommand{\@institute}{#1}}

\renewcommand\textsl{\textcolor{blue}}

\begin{document}
\title{Causal inference for data centric engineering}
\author[1]{Daniel J. Graham}
\affil[1]{Department of Civil Engineering, Imperial College London, London, SW7 2AZ, UK. \\ Email: \textcolor{Navy}{\url{d.j.graham@imperial.ac.uk}}} 
\date{}

\maketitle
\begin{abstract}
The paper reviews methods that seek to draw causal inference from observational data and demonstrates how they can be applied to empirical problems in engineering research. It presents a framework for causal identification based on the concept of potential outcomes and reviews core contemporary methods that can be used to estimate causal quantities. The paper has two aims: first, to provide a consolidated overview of the statistical literature on causal inference for the data centric engineering community; and second, to illustrate how causal concepts and methods can be applied. The latter aim is achieved through Monte Carlo simulations designed to replicate typical empirical problems encountered in engineering research. {\tt R} code for the simulations is made available for readers to run and adapt and citations are given to real world studies. Causal inference aims to quantify effects that occur due to explicit intervention (or `treatment') in non-experimental settings, typically for non-randomly assigned treatments. The paper argues that analyses of engineering interventions are often characterized by such conditions, and consequently, that causal inference has immediate and valuable applicability.
\end{abstract}

\section{Introduction}\label{Intro}

Engineering research often seeks answers to questions that involve cause-effect relationships. For example: what impacts do our engineering interventions have on ecosystems and climate change? Can the introduction of low emission zones achieve air quality improvement? Will congestion in cities decrease if we expand highway capacity? How will the adoption of connected and autonomous vehicle (CAV) technologies affect the safety and operations of highway networks? 

The motivation for this paper is the conviction that a correct response to such questions can only be realised when issues of {\em causality} are at the forefront, both of our conceptual and methodological research designs. Moreover, if we want to use public interventions to engineer future outcomes effectively, then a causal understanding of key relationships is essential.

At its core, causal inference is concerned with statistical quantification of the impacts that arise by engineering some change in a system. Causality is certainly an elusive concept to operationalise empirically. However, there is now a well established statistical literature on causal inference that is used routinely across the physical, biological and social sciences to understand the conditions and methods required to identify cause-effect relationships. Crucially, this causal inference framework is designed for use in non-experimental settings; that is, when the data available for empirical research are observed rather than generated through a controlled experiment, and this is a setting that often features strongly in data centric engineering (DCE) research.

The adoption of causal methods has been somewhat slower in fields of engineering than in other areas of science. In this paper, we provide a comprehensive review of causal inference concepts and methods and provide insights on how they can be applied in DCE research. In contrast to exiting reviews, emphasis is placed on models and methods for multivalued and continuous treatments, since engineering interventions often tend to be non-binary. Application to engineering problems is demonstrated through Monte Carlo simulation studies and {\tt R} code for the simulations is available for readers to run and adapt.

The paper is structured as follows. In section two we introduce preliminary concepts, set notation, and discuss the concept of causality in the context of DCE research. A framework for causal identification based on the concept of potential outcomes is then presented in section three. Methods for estimation of causal quantities are reviewed in sections four and five. Demonstrations of causal methods in DCE research problems are presented in section six and conducted via simulation. Conclusions are then drawn in the final section.     

\section{Preliminary concepts}\label{Prelim}

\subsection{Notation and structure of causal inference problems}

We will develop the concepts of causal inference in relation to random variable (or vector) $Z$, and assume that realizations, or observed data, $z$, are available to form inference for units $i$, $i=1,...,n$, in a sample of $n$ units from a population of interest.

As usual, the underlying aim of inference will be to infer properties of the parameters of the distribution of $Z$, $f_Z(z)$, which is potentially unknown and for which we postulate a statistical model. If we posit a parametric model, we will restrict the analytic form of $f_Z(z)$ to a suitable family and assume that it is determined by a finite number of real unknown parameters, $\theta=(\theta^1,...,\theta^d)$, that lie in parameter space $\Omega_{\theta}$. The class of densities for finite-dimensional parametric models can be written  
\[
\mathcal{P}=\left\{f_Z(z;\theta), z \in \mathcal{Z}, \theta  \in \Omega_{\theta} \subseteq \mathbb{R}^d\right\},
\]
where $\mathcal{Z}=\{z: f_Z(z;\theta)>0\}$ is the sample space in which the data lie and $f_Z(z;\theta)$ is the model function. 

Alternatively, we can apply nonparametric (or distribution free) inference and seek to model the data in the absence of a parametric representation. In this paper, we will interchange freely between parametric and nonparametric inference, typically setting up identification in a non parametric fashion and proposing parametric models for estimation. 

Given $Z$, and $f_Z(z)$, our discussion of causal inference will centre around three components:
\begin{enumerate}
\item The {\em Outcome}, or {\em Response}, of interest, which we denote $Y$.
\item The {\em Treatment} (or {\em Intervention}) to be studied, which we denote $D$. 
\item The {\em Covariates}, or measured pre-treatment characteristics, which we denote $X$. 
\end{enumerate}   

The typical set up, therefore, is one in which the data available for estimation take the form of a random vector $z_i = (y_i,d_i,x_i)$. Crucially, these data are assumed to be {\em observational}, that is generated passively via sampling from a population of units, rather than {\em experimental}, that is generated actively through an experiment conducted under controlled conditions. 

The key feature of inferential analyses with observational rather than experimental data is that the process that generated the observed data is unknown. This is important, as it is in fact aspects of the so called Data Generating Process (DGP), rather than mere associational inference about the parameters of $f_Z(z;\theta)$, that causal inference seeks to infer. Specifically, the key aim of causal inference is to estimate the {\em causal} effect on outcome $Y$, of a treatment (or intervention) $D$, and to do so in non-experimental settings typically for non-randomly assigned treatments. A central motivation for this paper is the belief that these conditions are pervasive in much engineering research. 

\subsection{Causality in Data Centric Engineering}

The key question that causal inference can help to answer is as follows: what effect(s) do our intervention(s) cause? Accordingly, we are interested in the causal effect that an engineering intervention (or a set of interventions) has on an outcome. We may want to know what the outcome would have been had the intervention not been applied, or if some different intervention been applied. 

For analytical purposes, we will view an intervention in a system as an observed realisation of a random variables whose manipulation can produce different outcomes. We refer to such random variables as `treatments', defined in the broadest sense to encompass any `regime' which can be manipulated to produce an effect. A treatment can even encapsulate an {\em exposure}, that is, some characteristic that units in the sample are exposed to rather than prescribed. Treatment variables can be binary, multivalued, or continuous. We provide in-depth coverage of models and methods for multivalued and continuous treatments because engineering interventions often tend to be non-binary, being characterised by length, volume, number, capacity and the like. 

Another important characteristic of engineering interventions is that they are typically non-randomly assigned. We intervene in systems to affect some changes, for example, to improve efficiency or effectiveness, and such interventions are rarely made without targeting, and are therefore non-random. This implies that `units' that receive engineering interventions will often tend to be in some sense different from those that do not. For analysis of cause-effect relationships, this creates a challenge because we need to somehow account for differences in baseline characteristics by treatment status in order to obtain valid causal inference.

\subsection{Data Generating Process, identification, and prediction}

Our representation of cause-effect relationships will make use of observed data, $z$, along with a model, $f_Z(z)$, derived from theory and scientific knowledge. Our aim will be to infer aspects of the Data Generating Process (DGP), that is, the process that we believe generated the data we observe. For instance, a fairly typical DGP for a causal problem could be represented as 
\begin{align*}
Y|D,X & \sim  \mathcal{N}(\beta_0 + \tau D + \beta_1 X, \sigma^2_Y)\\
D|X & \sim   \mathcal{B}(\text{expit}(\alpha_0+\alpha_1 X))\\
X & \sim \mathcal{N}(\mu_X,\sigma^2_X)
\end{align*}
Under this DGP, outcome $Y$ is determined by a Bernoulli distributed ($\mathcal{B}$) binary treatment $D$ and by normally distributed ($\mathcal{N}$) covariates $X$, while the treatment itself is assigned non-randomly in relation to $X$. The DGP is unknown to the analyst, and it is typically not the objective of causal inference to fully recover it, but rather to infer some aspect of it. For example, the parameter $\tau=\partial Y / \partial D$, which describes the causal effect of $D$ on $Y$, will be a common target of inference. 

In seeking to infer aspects of the DGP, a fundamental guiding principle of causal inference, addressed prior to and separate from estimation, is the need to achieve {\em identification}. The search for identification is concerned with determination of unique parameter values from sufficient data: essentially, we want to ensure that it is logically possible to uniquely determine the value of model parameters given an infinite number of observations.
\begin{Def}(Identification) a parameter $\theta$ for a family of distributions $\{f_Z(z;\theta): \theta \in \Omega_{\theta}\}$ is identifiable if distinct values of $\theta$ correspond to distinct pdfs or pmfs. That is, if $\theta \neq \theta'$, then $f_Z(z;\theta)$ is not the same function of $z$ as $f_Z(z;\theta')$
\end{Def} 
\begin{Def}(Observational equivalence) if $f_Z(z;\theta)=f_Z(z;\theta')$ then these two structures of the model are said to be observationally equivalent
\end{Def} 

The existence of multiple observationally equivalent structures implies that a model is not identified, because if observations from two distributions look identical, we cannot know whether the true value of the parameter is $\theta$ or $\theta'$. Non-identifiability precludes a causal interpretation of the DGP since there is more than one model structure that will generate the observed data.

The concept of identification is fundamental to causal inference because ultimately we are seeking to identify a unique set of parameters that capture causal relationships embedded in the DGP. That our focus is on {\em interventions} is important as a route to identification. The underlying philosophy of causal inference, as articulated by \citet{Holland:1986}, is to focus the inference problem on identifying the `effects of causes' rather than the deducing the cause of a given effect. 

It is worth noting that the aim of identification contrasts with that of prediction. Consider the model $y=f_X(x;\theta)$. For identification we aim to quantify uniquely some, but perhaps not all, parameters of our model. For example, we may be interested in estimation of a single parameter $\hat{\theta}_k = \partial y / \partial x_k$. For prediction, our  quantity of interest is $\hat{y}$. These two aims are closely related, but not the same. A perfectly identified parameter may, or may not, be particularly helpful in predicting $y$ in a multivariate model setting. Good prediction of $y$, however, can certainly be achieved in the absence of a perfectly identified model. 

The next two sections of the paper outline the defining characteristics of an approach to causal inference known as potential outcomes approach, and demonstrate how it can be used to infer cause-effect relationships. The focus in these sections is on core concepts and methods. We return to the issue of application in DCE research in section 6, where we demonstrate the use of causal methods in engineering related settings using simulation.     

\section{The potential outcomes framework for causal inference}\label{POF}

The framework we will use for causal inference is based on the concept of potential outcomes. It was first developed for binary treatments in a series of papers by Rubin \citep[e.g.][]{Rubin:1973a,Rubin:1973b,Rubin:1974,Rubin:1977,Rubin:1978}, although Rubin acknowledges inspiration for his approach in earlier work by \citet{Fisher:1935} and \citet{Neyman:1923}. A key feature of the potential outcomes framework is that it is essentially nonparametric. We will not specify the parametric form of the models used to represent causal inference, but will instead derive identification using only probability and expectation. Of course, an important implication of this is that the potential outcomes framework is applicable whatever form of parametric model we adopt.    

\subsection{Potential outcomes, causal estimands and identification}

First, we introduce the key features and challenges of our causal inference problem. We want to infer the effect that a treatment $D$ has on a defined outcome $Y$. The treatment in question could be binary, $\mathcal{D}\in\{0,1\}$, multivalued, $\mathcal{D}\equiv(d_0,d_1,...,d_m)$, or continuous, $\mathcal{D} \subseteq  \mathbb{R}$. 

For any level of treatment there is an associated {\em potential outcome}, defined as follows.
\begin{Def}(Potential Outcomes) For each unit $i$ we define $Y_i(d)$ as the potential outcome for unit $i$ when exposed to treatment $D_i=d$. The full set of potential outcomes for each unit is then $\mathcal{Y}_i=\{Y_i(d): d \in \mathcal{D}\}$. 
\end{Def}
Thus, for a binary treatment there are two potential outcomes for each unit, $Y_i(1)$ and $Y_i(0)$; for a multivalued treatment there are $m$ outcomes, $Y_i(d_0),Y_i(d_1),...,Y_i(d_m)$, and for a continuous treatment there are potentially infinite outcomes denoted in general by $Y_i(d)$ for $d$ in $\mathcal{D} \subseteq  \mathbb{R}$.

We will make use of potential outcomes to calculate the causal effect of $D$ on $Y$. There are three {\em causal estimands} that are of potential interest.  
\begin{Def}(Individual Causal Effect). The individual causal effect, 
\[\tau_i=Y_i(d)-Y_i(d_0), \ \ d \neq d_0,\] 
is the difference in outcomes for unit $i$ under treatment level $D_i=d$, relative to reference treatment level $D_i=d_0$. 
\end{Def}    

\begin{Def}(Average Potential Outcome (APO)). The average potential outcome, 
\[\mu(d)=\Expect[Y_i(d)],\] 
is the expected outcome had all unit in the population been treated at level $D_i=d$. 
\end{Def}

\begin{Def}(Average Treatment Effect (ATE)). The average treatment effect, 
\[\tau(d)=\mu(d)-\mu(d_0)=\Expect[Y_i(d)-Y_i(d_0)],\] 
is the difference in expected outcomes for all units in the population under treatment (or intervention) level $D_i=d$, relative to some other reference level of treatment $D_i=d_0$. 
\end{Def} 

Other causal estimands could include ATEs and APOs for sub-populations, conditional effects (e.g. conditional on covariates or on treatment status), quantile effects, and causal odds and risk ratios. In this paper the primary concern is with APOs and ATEs as defined above. Note that often the reference level of treatment is no treatment, or control, e.g. $d_0=0$; and in the case of binary treatments the ATE is defined $\tau(1)=\Expect[Y_i(1)-Y_i(0)]$, being the difference in expected outcomes under intervention (treatment) and nonintervention (control).   

Our ability to identify and consistently estimate causal estimands using observational data is determined fundamentally by two key challenges that characterise the practice of causal inference.              

{\bf Challenge 1: Missing Data}

The data available for estimation reveal only actual outcomes not potential outcomes. This means that of the full set of outcomes for unit $i$, $\mathcal{Y}_i$, we observe only one element, the {\em actual outcome}
\[Y_i=\sum_{d \in \mathcal{D}}I_{d}(D_i)Y_i(d),\]
where $I_{d}(D_i)$ is the indicator function for receiving treatment dose $d$, e.g. 
\begin{equation*}
I_{d}(D_i)\stackrel{}{=} \left\{
\begin{array}{ll} 1 & \text{if  } D_i = d. \\
0 & \text{if  } D_i \neq d
\end{array} \right..
\end{equation*}
Outcomes at all other levels, $d \neq D_i$, are unobserved and we refer to these as {\em counterfactual outcomes}. Thus, in the binary case we observe $Y_i=Y_i(1) I_{1}(D_i)+Y_i(0)(1-I_{1}(D_i))$, but we do not observe the joint density $f(Y_i(0),Y_i(1))$, since the two outcomes never occur together. 

\citet{Holland:1986} refers to this challenge as the `fundamental identification problem of causal inference'. An immediate implication is that we cannot observe ICEs, and in fact, ICEs are actually not identifiable because the available data do not provide enough information. For this reason, ICEs do not feature as targets of inference. 

A key insight of the potential outcomes approach is that causal effects can still be identified with missing potential outcomes if we focus on estimating average causal effects over the population rather than ICEs. In fact, if the treatment is {\em randomly assigned across the population}, then identification of quantities such as APOs and ATEs is possible because randomisation implies unconditional independence of treatment assignment and outcome. The argument is as follows.

\begin{Thm}\label{IUI}(Identification of the APO under unconditional independence). 
A random treatment assignment implies unconditional independence (notationally $ \independent$) of the form
\[Y_i(d) \independent  I_{d}(D_i)\]
for all $d \in \mathcal{D}$. If the potential outcomes are {\em unconditionally independent} of the treatment assignment, then APOs are identified. 
\end{Thm} 
 \begin{Proof}(Identification of the APO under unconditional independence). Unconditional independence allows for identification of the APO as follows 
\[\mu(d)=\Expect\left[Y_i(d))\right] =\Expect\left[Y_i(d)|D_i=d)\right]=\Expect\left[(Y_i|D_i=d)\right],\]
and thus we can estimate the conditional expectation of $Y$ given $D$ using observed data $(y,d)$. 
\end{Proof} 
Theorem (\ref{IUI}) justifies calculation of APOs and the ATE via sample means using
\begin{equation}
\widehat{\tau}(d)=\frac{\sum Y_i \cdot  I_{d}(D_i)}{\sum  I_{d}(D_i)}-\frac{\sum Y_i \cdot  I_{0}(D_i)}{\sum  I_{0}(D_i)}, 
\label{randATEcon}
\end{equation}
and consistency of estimation can be established by simply invoking a weak law of large numbers (WLLN): e.g. $n^{-1} \sum_i Y_i I_{d}(D_i) \overset{p}{\longrightarrow} \Expect\left[Y_i(d))\right]$.

However, while identification, and consequently consistency, can be established under a random treatment assignment, in reality treatments are often not randomly assigned. If so, theorem (\ref{IUI}) no longer applies and identification and consistency fail.   

{\bf Challenge 2: Non-random assignment and the problem of confounding}

This bring us to the second challenge in attaining causal identification with observational data: non-random assignment and confounding. We first define confounding. 
\begin{Def}(Confounding) 
A treatment assignment is said to be confounded if characteristics of the data generating process create unconditional dependence between potential outcomes and treatment assignment such that
\[Y_i(d) \not\independent  I_{d}(D_i)\]
for $d \in \mathcal{D}$. Confounding is present when a non-random treatment assignment mechanism induces dependency with a set of covariates $X_i$, which are themselves important in determining outcome $Y_i$.
\end{Def} 
A confounder is, therefore, a random variable that simultaneously determines both the outcome of interest and (non-random) assignment to treatment. Confounding can manifest dynamically, with past outcomes or treatments serving as baseline confounders (e.g. $X$ can contain lagged values of $D$ and $Y$). 

Figure \ref{C2F1} below shows a graphical comparison of randomised (unconfounded) and non-randomised (confounded) treatment assignments.
\begin{figure}[htp]
\centering 
\includegraphics[height=5.0cm]{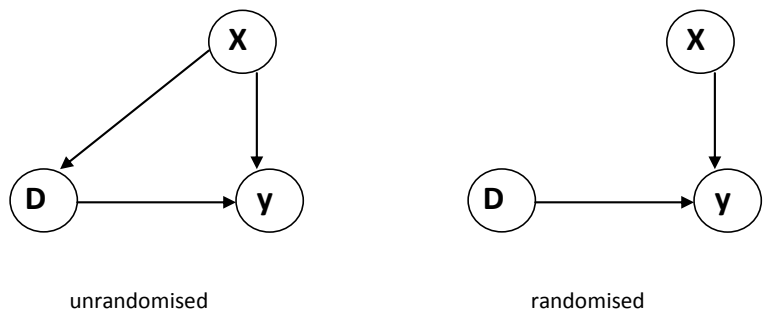}
{\caption{Directed Acyclic Graph of observational data with randomisation and non-randomisation of treatment assignment\label{C2F1}}}
\end{figure}

Under a random assignment unit characteristics $X_i$ have no influence on the treatment received (i.e. on $D_i$), and consequently there are no {\em systematic} differences between units receiving different levels of the treatment. As explained above, this property of unconditional independence allows us to treat unobserved potential outcomes much like data that are missing at random and form consistent estimators of APOs and ATEs via sample means.

Under non-randomisation, however, assignment of the treatment depends on a set of covariates $X_i$ which are themselves important in determining outcome $Y_i$. Thus, some part of the association between the treatment and the outcome could be attributed to $X_i$ rather than $D_i$. Under these circumstance we refer to $X_i$ as {\em confounders} and note that simple comparisons of mean responses across different treatment groups will not in general reveal a `causal' effect because mean outcomes across treated and control units can differ regardless of treatment status. 

\subsection{Identification of causal effects in the presence of confounding via the potential outcomes framework}

While the two challenges defined above have consequence for causal identification, consistent estimates of APOs and ATEs can still be obtained under the potential outcome framework. In this subsection of the paper, we define the conditions under which causal identification can be achieved in the presence of missing data and confounding. There are some important differences between the identification conditions for binary and multivalued / continuous treatments, so we will describe both.       

Within the potential outcomes framework there are three key assumptions required for valid APO and ATE identification in the presence of confounding. These are as follows. 
\begin{Assum}(Conditional independence). The potential outcomes for unit $i$ must be conditionally independent of the treatment assignment given a (sufficient) set of observed pre-treatment covariates $X_i$. For binary treatments the assumption requires that
\begin{equation}
Y_i(0),Y_i(1)) \independent I_{1}(D_i)|X_i,
\label{CI1}
\end{equation}
and for multivalued or continuous treatments \citet{Imbens:2000} and \citet{Hirano/Imbens:2004} introduce the concept of weak conditional independence which can be stated as
\begin{equation}
Y_i(d) \independent  I_{d}(D_i)|X_i \ \ {\rm for \ all} \ \ d \in \mathcal{D}.
\label{CI2}
\end{equation} 
\end{Assum}
The key difference between the binary and non binary assumptions is that in the latter conditional independence is required to hold for each value of the treatment (i.e. pairwise), but not joint independence of all potential outcomes.

The conditional independence assumption essentially requires that, conditional on some set of pre-treatment covariates, assignment to treatment does not depend on the outcome. If $X_i$ is sufficient for this to hold then we can in effect mimic, with confounded observational data, the assignment that would occur in a randomised control trial in which the treatment is allocated independently of pre-treatment characteristics.
 
\begin{Assum}(Common support). The support of the conditional distribution of $X_i$ given a particular treatment status should overlap with that of $X_i$ given any other treatment status. For binary treatments this requires that the probability of assignment to the treatment lies strictly between zero and one
\begin{equation}
0<\P(I_{1}(D_i)=1|X_i=x)<1, \ \forall \ x.
\label{C2As3}
\end{equation}
For multivalued or continuous treatments we require common support by treatment status in the covariate distributions within some region of dose $\mathcal{C}\subseteq \mathcal{D}$. A sufficient condition is that for any subset of $\mathcal{C}$, say $\mathcal{A}\subseteq \mathcal{C}$, 
\begin{equation}
\P(D_i \in \mathcal{A}|X_i=x)>0, \ \forall \ x
\label{C2E7}
\end{equation}
\end{Assum}
The intuition behind the common support, or overlap, assumption is that if some sub-populations observed in $X_i$ have zero probability of receiving (or not receiving) a treatment, then it does not make sense in these cases to talk of a treatment effect since the counterfactual does not exist in the observed data.

\begin{Assum}(Stable unit treatment values). The relationship between observed and potential outcomes must comply with the Stable Unit Treatment Value Assumption (SUTVA)\citep[e.g.][]{Rubin:1978,Rubin:1980,Rubin:1986,Rubin:1990}, which requires that the observed response under a given treatment allocation is equivalent to the potential response under that treatment allocation. For binary treatments we require that 
\begin{equation}
Y_i=I_{1}(D_i)Y_i(1)+ (1-I_{1}(D_i))Y_i(0)
\label{C2As1}
\end{equation}
for all $i=1,...,N$. For multivalued or continuous treatments we require
\begin{equation}
Y_i \equiv I_{d}(D_i)Y_i(d)
\label{C2E8}
\end{equation}
for all $d \in \mathcal{D}$, for all $Y_i (d) \in \mathcal{Y}_i$, and for $i=1,...,N$. 
\end{Assum}
The SUTVA effectively imposes two conditions: first, that the outcome for each unit be independent of the treatment status of other units, or in other words, there should be no interference in treatment effects across units \citep{Cox:1958}; and second, that there are no different versions of the treatment. The no interference assumption is generally satisfied when the units are physically distinct and have no means of contact. Violations of the assumption can occur when proximity of units allows for contact and this presents a particular concern for some engineering applications, particuraly those that are network based \citep[e.g.][]{Graham/et/al:2013}.

The three assumptions defined above, which are together referred to by \citet{Rosenbaum/Rubin:1983b} as {\it strong ignorability}, permit identification of APOs and ATEs for non randomly assigned treatments. This is stated formally in the following theorem.  
\begin{Thm}\label{ISI}(Identification of the APO and ATE under strong ignorability). Under strong ignorability APOs and ATEs for non randomly assigned treatments can be identified by conditioning on $X$ and integrating over the covariate distributions to capture the marginal causal effect. 
\begin{align*}
\tau(d) &=\Expect[Y(d)-Y(0)]=\int \Expect[Y(d)-Y(0)|X]dF(X)\\
&=\int \Expect\left[Y|X,I_d(D)\right]-\Expect\left[Y|X,I_0(D)\right]dF(X)
\end{align*}
\end{Thm}
Note that the ATE is defined as an expectation over covariates $X$. If we do not take this expectation, but instead simply use the integrand, we obtain an estimate of the causal effect of $D$ within strata of $X$. In other words, we get the {\it conditional treatment effect}, that is the ATE for units with characteristics $X=x$. By integrating $X$ out of this distribution we get the average causal intervention distribution. 

The justification for theorem (\ref{ISI}), in the case of binary treatments, is as follows. 
\begin{Proof}(Identification of a binary ATE under strong ignorability)  
\begin{subequations}
\label{iden}
\begin{align}
\tau(1) =& \Expect_i\left[Y_i(1)-Y_i(0)\right]\\
=& \Expect_X\left[\Expect_i(Y_i(1)|X_i=x)-\Expect_i(Y_i(0)|X_i=x)\right]\\
=&\Expect_X\left[\Expect_i(Y_i(1)|X_i=x, I_{1}(D_i)=1)-\Expect_i(Y_i(0)|X_i=x, I_{1}(D_i)=0)\right]\\
=& \Expect_X\left[\Expect_i(Y_i|X_i=x, I_{1}(D_i)=1)-\Expect_i(Y_i|X_i=x,I_{1}(D_i)=0)\right].
\end{align}
\end{subequations}
\end{Proof}
The law of iterated expectations gets us from (\ref{iden}a) to (\ref{iden}b), conditional independence justifies the equality of (\ref{iden}b) and (\ref{iden}c), the SUTVA allows the substitution of observed for potential outcomes to give (\ref{iden}d), and overlap ensures that the population ATE in (\ref{iden}d) is estimable since there are units in both the treated and untreated groups. 

For continuous or multivalued treatments, identification of the APO, $\mu(d)=\Expect[Y_i(d)]$, under a given dose $D=d$, or the dose-response function, can be demonstrated as follows.
\begin{Proof}(Identification of a multivalued or continuous APO under strong ignorability)  
\begin{subequations}
\label{idenMVC}
\begin{align}
\mu(d)= &\Expect[Y_i(d)]\\
= &\Expect_X \left[\Expect(Y_i(d)|X_i) \right]\\
= &\Expect_{X} \left[\Expect(Y_i(d)|I_{d}(D_i),X_i) \right]\\
= &\Expect_{X} \left[\Expect(Y_i|I_{d}(D_i),X_i) \right], \label{APO}
\end{align} 
\end{subequations}
\end{Proof}
where (\ref{idenMVC}c) follows from conditional independence, (\ref{idenMVC}d) from the SUTVA, and the overlap assumption ensures that (\ref{idenMVC}d) is estimable since there are comparable units across treatment levels. 

These proofs therefore demonstrate causal identification for average causal effects for non-randomly assigned treatments, in the sense that they show we are able to uniquely determine average causal quantities with observational data. In the next two sections, we turn to methods that draw on these principles to estimate causal parameters. We first review methods that assume ignorability has been met, and then consider approaches that can be used when ignorability is thought to be violated.   

\section{Methods for treatment effect estimation under ignorability}\label{CMUI} 

Under strong ignorability, estimation of APOs and ATEs can be achieved via outcome regression (OR) models, Propensity Score (PS) models, and mixed or Doubly Robust (DR) models. To emphasise how these approaches differ, we adopt the notation of \citet{Tsiatis/Davidian:2007} and define joint densities of the observed data in the form
\begin{equation}\label{TD2007}
f_Z(z)=f_{Y|D,X}(y|d,x)f_{D|X}(d|x)f_X(x).
\end{equation}
OR models focus on $f_{Y|D,X}(y|d,x)$, PS models on $f_{D|X}(d|x)$ and DR models use both. We now explain approach in turn.    
 
\subsection{Outcome Regression (OR) Models}
OR approaches are widely used across scientific disciplines and will likely be familiar to most, so they can be covered here briefly and in outline. In the causal inference setting, and with reference to (\ref{TD2007}), the OR model leaves $f_{D|X}(d|x)$ and $f_{X}(x)$ unspecified, and explicitly models the conditional expectation function $\Expect[Y_i|D_i,X_i]$. The focus on this conditional expectation is motivated by (\ref{iden}) and (\ref{idenMVC}), which show that identification can be achieved by averaging over this function.   

The conditional expectation, or mean response, could be specified as a linear regression model, a Generalized Linear Model (GLM), a Generalized Linear Mixed Model (GLMM), or some parametric or semiparametric variant thereof. If the OR model, denoted $\Psi^{-1}\{m(D_i,X_i;\xi)\}$, for known link function $\Psi$, is correctly specified for the conditional expectation, then consistent estimation of ATE can be achieved using
\[\widehat{\tau}_{OR}(1)=\frac{1}{n}\sum_{i=1}^n\left[\Psi^{-1} \left\{m(1,X_i; \widehat{\beta})\right\}-\Psi^{-1} \left\{m(0,X_i; \widehat{\beta})\right\} \right],\]   
in the binary case and
\[\widehat{\tau}_{OR}(d)=\frac{1}{n}\sum_{i=1}^n\left[\Psi^{-1} \left\{m(d,X_i; \widehat{\beta})\right\}-\Psi^{-1} \left\{m(0,X_i; \widehat{\beta})\right\} \right],\] 
in the case of continuous treatments. 

In other words, simply taking the mean of the predicted values of the OR model, with $D$ held fixed at some value $d$, gives a consistent estimate of the APO $\mu(d)= \Expect[Y_i(d)]$. 

Note, however, that consistency of OR based adjustment requires that the observed $X$ covariate vector adequately represents {\em all} sources of confounding, and thus is sufficient to invoke conditional independence. In practice, the true DGP is not known, and there is in fact no way to objectively test whether the CIA has been satisfied or not. Theory provides a basis upon which sources of confounding (and simultaneity) can be hypothesised. If our assumed DGP indicates the existence of confounders that we cannot measure, then we cannot achieve identification via OR adjustment with measured covariates alone. 

If we believe that unmeasured confounding is present then we must turn to methods that offer robustness to this issue. Several such approaches are reviewed in section \ref{ENITA} of the paper. However, there is another widely used class of methods that exploit the structure of longitudinal (or panel) data to represent unobserved confounding. We briefly review these methods here as they represent a straightforward variant of the standard OR approach. 

\subsubsection{Longitudinal / Panel Data OR Models}\label{LPDORM}

A panel, or longitudinal, data structure comprises $N$ units (or subjects), $i=1,...,N$, each of which has $n_i$ measurements made over times $t$, $t=1,...,n_i$, giving a total of $n=\sum_{i=1}^{N}n_i$ sample observations. The data thus capture variation over both cross-sectional units and time.

There are many advantages of panel data including the potential for increased precision of estimation, inclusion of dynamics, and modelling of unobserved heterogeneity \citep[see][]{Wooldridge:2010}. In the context of causal inference, however, panel OR models are especially useful because they provide opportunities to adjust for unobserved unit level confounding.

Consider, for instance, a CIA for a longitudinal data structure of the form
\begin{equation}\label{PDCIA}
Y_{it}(d) \independent D_{it}|X_{it},W_{i} \ \text{for all} \ d \in \mathcal{D}.
\end{equation}

This CIA states that independence of response and treatments assignment is conditional on a set of {\em time-varying} covariates, $X_{it}$, and on a set of {\em time-invariant} covariates, $W_{i}$. Furthermore, suppose that we do not observe $W_{i}$. Then the APO estimate from an OR model based purely on observables,
\[\widehat{\mu}(d)=\frac{1}{n}\sum_{i=1}^n\Psi^{-1} \left\{m(d,X_{it}; \widehat{\beta})\right\}, \] 
will not produce a consistent or unbiased estimate of $\mu(d)$ because the CIA does not hold and thus the argument for identification presented in (\ref{idenMVC}) fails.  

Under these conditions, we can utilise a panel data model to indirectly represent $W_i$, and invoke the full CIA (e.g. \ref{PDCIA}) conditional on our indirect representation. Typically, this is done by introducing a set of parameters, $\alpha_i$, into the panel model, which vary by cross-sectional unit and effectively adjust for unobserved time-invariant confounding. The properties of $\alpha_{i}$ determine the ability of a panel specification to correct for confounding, specifically they have to allow for correlation between the treatment and the unobserved effects. Five commonly used panel model specifications are as follows. 
\begin{enumerate}
\item[{\bf 1.}] {\bf Pooled Model (POLS)} -  $\alpha_i=\alpha$ is assumed constant for all cross-sectional units implying that $\Cov(d_{it},\alpha_i)=0$. 
\item[{\bf 2.}] {\bf Random Effects (RE)} - the $\alpha_i$ terms are given an error structure: $\alpha_i \sim IID(0, \sigma^{2}_{\alpha})$, which accommodates within group error correlation but assumes they are uncorrelated with the covariates, and thus $\Cov(d_{it},\alpha_i)=0$. 
\item[{\bf 3.}] {\bf Fixed Effects (FE)} -  $\alpha_i$ is treated as a fixed (e.g. no distribution) unobserved variable that is potentially correlated with the covariates, and thus $\Cov(d_{it},\alpha_i)\neq 0$ . 
\item[{\bf 4.}] {\bf First-Differences (FD)} - the $\alpha_i$ terms are removed from the model by differencing the data. First differencing allows for $\Cov(d_{it},\alpha_i)\neq 0$. 
\item[{\bf 5.}] {\bf Correlated Random Effects (CRE)} - $\alpha_i= \bar{D_i}^{\sf{T}}\xi + \omega_{i}$, where $\bar{D}_{i,1} = n_i^{-1}\sum_{t=1}^{n_i} D_{it,1}$) with parameter vector $\xi$ and random component $\omega_{i}\sim \mathcal{N}(0,\sigma^2_{\omega})$; and because $\Expect[\alpha_i|D_{it}]=\bar{d_i}^{\sf{T}}\xi$, $\Cov(d_{it},\alpha_i)\neq 0$.
\end{enumerate}

The table below summarises the properties of the panels models in relation to estimation of $\mu(d)$.

\begin{table}[htbp]
  \centering
  \caption{Properties of panels models for estimation of $\mu(d)$}
    \begin{tabular}{lcccc}
    \hline
              & {\em no unmeasured}      & {\em no unmeasured}    & {\em unmeasured} & {\em unmeasured} \\
          & {\em confounding}      & {\em confounding}    & {\em time-invariant} & {\em time-varying} \\
    {\em Model} & {$\alpha_i=\alpha$} & {\em $\alpha_i\sim N(0,\sigma^2_{\alpha})$} & {\em confounding} & {\em confounding} \\
    \hline
    {Pooled Model}  & {\textcolor{mygreen}{$\checkmark$}}  & {\textcolor{mygreen}{$\checkmark$}} & {\bf \textcolor{myred}{$\times$}} & {\bf \textcolor{myred}{$\times$}} \\
        {Random Effects} & {\textcolor{mygreen}{$\checkmark$}}  & {\textcolor{mygreen}{$\checkmark$}} & {\bf \textcolor{myred}{$\times$}} & {\bf \textcolor{myred}{$\times$}} \\
    {Fixed Effects} & {\textcolor{mygreen}{$\checkmark$}}  & {\textcolor{mygreen}{$\checkmark$}} & {\textcolor{mygreen}{$\checkmark$}} & {\bf \textcolor{myred}{$\times$}} \\
 {First Differences} & {\textcolor{mygreen}{$\checkmark$}}  & {\textcolor{mygreen}{$\checkmark$}} & {\textcolor{mygreen}{$\checkmark$}} & {\bf \textcolor{myred}{$\times$}} \\
    {Correlated Random Effects} & {\textcolor{mygreen}{$\checkmark$}}  & {\textcolor{mygreen}{$\checkmark$}} & {\textcolor{mygreen}{$\checkmark$}} & {\bf \textcolor{myred}{$\times$}} \\
    \hline
    \end{tabular}%
  \label{tab:addlabel}%
  \footnotesize
  \\
Key: {\textcolor{mygreen}{$\checkmark$}} - consistent for $\mu(d)$ (identified), {\bf \textcolor{myred}{$\times$}} - inconsistent for $\mu(d)$ (not identified)
\end{table}%

\normalsize

With no unobserved confounding all models are consistent for $\mu(d)$, but some may be more efficient if there is time-invariant heterogeneity (e.g. POLS and RE). The POLS and RE models are inconsistent for $\mu(d)$ in the presence of unobserved time-invariant confounding, while the FD, FE and CRE models are consistent. No panel specification is robust to time-varying confounding.  

Thus, panel models that allow for $\Cov(d_{it},\alpha_i)\neq 0$ offer a way of adjusting for unit level time-invariant characteristics that we do not observe, but that we believe induce confounding. It is this added robustness to the CIA assumption that lies behind the widespread use of these methods, and they are to be recommended over cross-sectional specifications when the available data allow them to be used.            

\subsection{Propensity Score (PS) Models}

The focus of OR models is on the relationship between outcome and treatment conditional on covariates. Alternatively, we can model causal relationships by studying the relationship between the treatment assignment and the covariates. Thus, with reference to (\ref{TD2007}) above, we leave $f_{Y|X}(y|x)$ and $f_X(x)$ unspecified, but assume a model for $f_{D|X}(d|x)$. This model is used to form {\em Propensity Scores} (PS), which measure the probability of assignment to treatment given the set of observed pre-treatment covariates \citep[introductions to the main ideas underpinning the PS are given in][]{Joffe/Rosenbaum:1999,Rosenbaum:1999,Rubin:2006}.

\begin{Def}
(Propensity Score). The propensity score measures the conditional probability of assignment to treatment given pre-treatment covariates $X_i$. For binary treatment the PS is defined
\[\pi\left(D_i=1|X_i\right)=\P(I_{1}(D_i)=1|X_i=x)\]
and for multivalued or continuous treatments 
\[\pi\left(D_i|X_i\right)=f_{D|X}(d|x_i)\]
\end{Def}

Before reviewing the main estimation methods based on the PS, we first set out the key identification issues. 

\subsubsection{Identification via Propensity Scores}

An important result, due to \citet{Rosenbaum/Rubin:1983b}, is that the CIA (i.e. equations \ref{CI1} and \ref{CI2}) can be restated by replacing the covariate vector $X_i$ with the scalar PS. \citet{Rosenbaum/Rubin:1983b} proved this result in the case of binary treatments, and \citet{Imbens:2000} and \citet{Hirano/Imbens:2004} generalize the PS to cover the the case of multivalued and continuous treatments. The reason the CIA can be established conditional on the PS, rather than full covariate vector, is because the PS has a {\em balancing property}, in the sense that it balances the distribution of the observed covariates within strata of of the sample that have the same PS. Balancing does not eliminate heterogeneity in $X_i$, but renders it such that on average units with the same PS can be treated as observationally equivalent. The logic is as follows. 
\begin{Lemma}\label{BP}(Balancing of pre-treatment covariates given the propensity score). If $\pi\left(D_i|X_i\right)$ is the propensity score, then
\begin{equation*}
I_{d}(D_i) \independent X_i |\pi(d|X_i).
\end{equation*}
This follows because $\pi(d|X_i)$ is a function of $X_i$ and so conditioning on $X_i$ adds no additional information, hence
\[\Expect\left[I_{d}(D_i )|X_i,\pi(d|X_i)\right]=\Expect\left[I_{d}(D_i )|\pi(d|X_i)\right].\]
\end{Lemma}
Proofs for lemma \ref{BP} are provided in the appendix. 

From lemma (\ref{BP}) it follows that conditional independence can be stated using the PS rather than $X$.
\begin{Corol}\label{Cor1}(Conditional independence given the propensity score). Given balancing, we can establish conditional independence on the scalar PS rather than the potentially high-dimensional covariate vector $X_i$, e.g. 
\[Y_i(d) \independent I_{d}(D_i)|\pi(d|X_i).\] 
\end{Corol}

Bringing all of the above together, if strong ignorability holds, and the model is correctly specified, the PS effectively adjusts for confounding and can thus be used for causal identification. 

\begin{Thm}(Identification given the propensity score). Suppose that assignment to the treatment is ignorable given pre-treatment characteristics $X_i$ and define 
\[\beta(d,\pi(d|X_i))=\Expect[Y_i(d)|I_{d}(D_i),\pi(d|X_i)],\]
as the conditional mean of the outcome given the treatment level $D=d$ and the the PS. If strong ikgnorability holds then the expectation of $\beta(d,\pi(d|X_i))$ over $X_{i}$ is an unbiased estimator of the APO
\[\Expect_{X_i}\left[\beta(d,\pi(d|X_i))\right]=\Expect_{X_i}\left[\Expect[Y(d)|X_i]\right]=\Expect[Y_i(d)]=\mu(d).\] 
\end{Thm}
Proofs for this result are provided in the appendix.

\subsubsection{Estimation via Propensity Scores} 

Given that identification can be achieved via the PS, we can construct PS based estimators for causal quantities. PSs are not observed but are calculated by estimating the relationship between $D$ and $X$ using a regression model 
\[\Expect[D_i|X_i]=\Psi^{-1}\{m(D_i,X_i; \alpha)\}\] 
for link function $\Psi$, regression function $m()$, and unknown parameter vector $\alpha$. The estimated parameters of this model are then used to compute propensity scores, $\pi(D_i|X_i; \widehat{\alpha})$.  

The estimated PS, $\pi(D_i|X_i; \widehat{\alpha})$, can then be used to form a number of different nonparametric and semiparametric estimators. A key advantage in using the PS is that it avoids the need to condition on a potentially high dimensional covariate vector, and it is this dimensions reducting property that allows for effective implementation of flexible estimators. Another advantage of the PS is that it is highly effective in isolating the region of common support, a task that is difficult using multiple covariates \citep[for discussion see][]{Joffe/Rosenbaum:1999}.

Below we briefly describe the most commonly used PS estimators which are realised via weighting, regression, stratification and matching. For further details see \citet{Imbens/Wooldridge:2009,Imbens/Rubin:2015,Graham/et/al:2014}.

{\em Inverse propensity score weighting}

Inverse PS weighting (IPW) provides a very simple estimator which is based on the notion that the APO, $\Expect[Y_i(d)]$, can be estimated using 
\begin{equation}
\widehat{\mu}_{IPW}(d)=\frac{1}{n}\sum_{i=1}^n \left[\frac{I_{d}(D_i) \cdot Y_i}{\pi(D_i|X_i;\widehat{\alpha})}\right],
\label{eq6}
\end{equation}
which by the WLLN is consistent if the PS model is correctly specified,e.g.
\[\frac{1}{n}\sum_{i=1}^n \left[\frac{I_{d}(D_i) \cdot Y_i}{\pi(D_i|X_i;\widehat{\alpha})}\right] \overset{p}{\longrightarrow} \Expect\left[\frac{I_{d}(D_i) \cdot Y_i}{\pi(D_i|X_i;\widehat{\alpha})}\right]\]
and by the central limit theorem, $\sqrt{N}$ asymptotically normally distributed \citep[e.g.][]{Rosenbaum:1987,Joffe/Rosenbaum:1999,Hirano/et/al:2003}. A proof for identification of $\mu(d)$ via IPW is provided in the Appendix.

From (\ref{eq6}) IPW can be use to estimate an ATE for binary treatments of the form
\begin{equation}
\widehat{\tau}_{IPW}(1)=\frac{1}{n}\sum_{i=1}^n \left[\frac{I_{1}(D_i) \cdot Y_i}{\pi(D_i=1|X_i;\widehat{\alpha})} - \frac{(1-I_{1}(D_i)) \cdot Y_i}{1-\pi(D_i=1|X_i;\widehat{\alpha})}\right],
\label{IPWB}
\end{equation}
and for continuous and multivalued treatments of the form
\begin{equation}
\widehat{\tau}_{IPW}(d)=\frac{1}{n}\sum_{i=1}^n \left[\frac{I_{d}(D_i) \cdot Y_i}{\pi(D_i=d|X_i;\widehat{\alpha})} - \frac{(1-I_{d}(D_i)) \cdot Y_i}{1-\pi(D_i=d|X_i;\widehat{\alpha})}\right],
\label{IPWD}
\end{equation}

The IPW estimator originated in \citet{Horovitz/Thompson:1952}. The principle it invokes revolves around creation of a {\em pseudo-sample} to simulate random assignment, and is achieved by using the conditional probabilities captured in the PS values. Under randomisation, assignment to treatment has uniform probability across units in the sample. The PS measures deviation from this uniform probability due to baseline characteristics, and accordingly we can use PS weighting to effectively alter the representation of units within the sample to mimic an assignment that is, for all intents and purposes, {\em as good as random}. This same principle is used in so called Marginal Structural Models (MSMs) \citep[see][]{Robins:1998b,Robins:1999,Robins:1999b,Robins/et/al:2000a}, which make use of inverse probability weighting in a slightly different form, but are otherwise conceptually very similar. MSMs, along with G-computation \citep[e.g.][]{Robins:1986} are useful in estimation of dynamic causal relationships, in which time varying covariates are present that act simultaneously as confounders and as intermediate variables.  

Note that the consistency of ATE estimation under IPW relies on the PS model being correctly specified. That is, the estimated PSs must provide an accurate measure of the conditional probability of assignment to treatment.   

{\em Regression on the propensity score (PSR)} 

Since the CIA can be established conditional on the PS rather than covariate vector $X_i$, PSs can be substituted into a regression model in place of $X_i$. Thus, adapting the OR model discussed above, we can estimate an ATE for treatment dose $D_i=d$ using
\begin{equation}
\widehat{\tau}_{PSR}(d)=\frac{1}{n}\sum_{i=1}^n\left[\Psi^{-1} \left\{m(d,\pi(d|X_i;\widehat{\alpha}); \widehat{\gamma})\right\}-\Psi^{-1} \left\{m(0,\pi(0|X_i;\widehat{\alpha}); \widehat{\gamma})\right\} \right]
\end{equation}
where $\gamma$ is a set of parameters to be estimated and $\widehat{\alpha}$ are parameters of the PS model estimated in a prior step. The logic underpinning PSR identification simply follows from the SUTVA and the balancing property of the PS, and is as follows 
\[\mu(d) = \Expect[Y_i(d)] = \Expect_X\left[\Expect(Y_i(d)|X_i)\right] = \Expect_X\left[\Expect(Y_i(d)|\pi(d|X_i;\alpha))\right] =\Expect_X\left[\Expect(Y_i|I_{d}(D_i),\pi(d|X_i;\alpha))\right].\]

A potential advantage of PSR is that by reducing the dimensionality of the model, that is by using a scalar PS in place of covariate vector $X_i$, estimation via parametric polynomial models or semiparametric spline models could produce a good approximation to the conditional expectation. However, a potential disadvantage, noted by \citet{Imbens/Wooldridge:2009}, is that since the PS itself is purely a statistical quantity, with no substantive meaning, interpretation of the regression results, and thus evidence on the nature of confounding, could be somewhat opaque. 

{\em Propensity score stratification (Blocking)}

PS stratification is another approach, suggested by \citet{Rosenbaum/Rubin:1983b}, and involves using the PS, or discretised values of this score, to partition the sample into strata within which there is little variation in the PS. Differences between treated and control groups are then analysed within each stratum. By maintaining the underlying assumption that the PS is roughly constant within each stratum, the data are treated as if coming from a randomised design. Thus, within stratum $j$, the ATE for a binary treatment is simply calculated as the difference in the mean outcomes for treated and control units
\[\hat{\tau}_{PSS_j}=\bar{Y}_{j1}-\bar{Y}_{j0},\]
and taking a weighted average of the within stratum estimates gives the sample ATE
\begin{equation}
\hat{\tau}_{PSS}=\sum_{j=1}^J \hat{\tau}_j \cdot \left( \frac{n_{j0}+n_{j1}}{n}\right).
\end{equation}
\citet{Imbens:2004} notes that the blocking estimator can be interpreted as a rough form of nonparametric regression which uses a step function on fixed jump points to approximate the unknown function. A key issue is how to adequately define the blocks. Following results by \citet{Cochran:1968} showing small bias for five blocks, quintiles have often been used in applied work \citep[e.g.]{Rosenbaum/Rubin:1983a,Rosenbaum/Rubin:1983b,Rosenbaum/Rubin:1984,Dehejia/Wahba:1999,Becker/Ichino:2002}, with procedures to check for adequate covariate balance also being applied. Alternatively, machine learning algorithms could be applied to define the blocks in a data driven fashion.    

{\em Matching on the propensity score}   
    
A procedure that is similar in principle to blocking is that of matching on the PS. The principle behind matching methods is to find units for the treated and control sub-samples that are as similar to each other as possible. The matching estimator has been used extensively in the literature \citep[see for example][]{Rosenbaum:1989,Rosenbaum:1995a,Rosenbaum:2002,Gu/Rosenbaum:1993,Rubin:1973a,Rubin:1973b,Rubin:1979,Rubin/Thomas:1992a,Rubin/Thomas:1992b,Rubin/Thomas:1996,Rubin/Thomas:2000,Dehejia/Wahba:1999,Becker/Ichino:2002,Abadie/Imbens:2002}. In its simplest form, this means fitting an algorithm to identify treated and control units that are similar across all covariates. In many applications, a one-on-one mapping of treatment to control units is attempted, other approaches match a small number of `neighbours' to approximate the unobserved outcome for a given unit. This gives the simple matching estimator
\begin{equation}
\hat{\tau}_M=\frac{1}{n}\sum^n_{i=1}(\hat{Y}_i(1)-\hat{Y}_i(0)),
\end{equation}    
where 
\begin{equation*}
\hat{Y}_i(1)\stackrel{}{=} \left\{
\begin{array}{ll} Y_i & \text{if  } I_{1}(D_i)=1 \\
\frac{1}{M}\sum_{j \in \mathcal{J}_{M(i)}} Y_j & \text{if  } I_{1}(D_i)=0
\end{array} \right., 
\end{equation*}
and
\begin{equation*}
\hat{Y}_i(0)\stackrel{}{=} \left\{
\begin{array}{ll} Y_i & \text{if  } I_{1}(D_i)=0 \\
\frac{1}{M}\sum_{j \in \mathcal{J}_{M(i)}} Y_j & \text{if  } I_{1}(D_i)=1
\end{array} \right. ,
\end{equation*}
where $\mathcal{J}_{M(i)}$ is the matched set of $1,...,M$ control units for unit $i$.

The properties of simple matching estimators have been studied extensively by \citet{Abadie/Imbens:2002} and \citet{Abadie/Imbens:2006}. They show that if the dimensions of the continuous covariates is greater than two and if the matching is not exact, then the matching estimator is not consistent with bias due to matching discrepancies of order $O(n^{-1/k})$, where $K$ is the dimension of the continuous covariates. \citet{Abadie/Imbens:2006} also show that matching estimators are generally not efficient. However, matching can work well in certain settings, and particularly when combined with other procedures such as regression.  

\subsection{Doubly Robust (DR) Models}

So far we have reviewed two classes of causal method based on OR and PS models. Consistency of causal estimation under either of these two classes relies on model correct specification, and if the models are misspecified in some way, then it can no longer be guaranteed. It is this issue of susceptibility to misspecification that has given rise to a class of doubly robust (DR) models, which combine OR and PS models to invoke an enhanced property of robustness.  

With reference to (\ref{TD2007}) above, under a DR approach, we leave $f(x)$ unspecified and assume a model for both $f_{Y|X}(y|x)$ and $f_{D|X}(d|x)$, and form an estimator that combines both OR and PS models. The key feature of DR estimators is that APO and ATE estimates are consistent and asymptotically normal when either the OR or the PS model are correctly specified, but we do not require both models to be correct \citep[e.g.][]{Robins:2000,Robins/et/al:2000c,Robins/Rotnitzky:2001,VanDerLaan/Robins:2003,Lunceford/Davidian:2004,Bang/Robins:2005,Kang/Schafer:2007}.  

The motivation for doubly-robust estimation, therefore, is that the analyst effectively has two chances at getting a model specification right. In practice, double robustness is typically achieved by weighting or augmenting the OR model with an inverse PS covariate, often referred to as a `clever covariate'. A statement of the DR property is as follows. 
\begin{Thm} (Double-robustness) An APO or ATE estimator, formed by combining an OR model $\Psi^{-1}\{m(X_i,D_i;\beta)\}$ and a PS model $\hat{\pi}(D_i|X_i;\hat{\alpha})$, is doubly-robust if the estimator is consistent when either $\Psi^{-1}\{m(X_i,D_i;\beta)\} $ is correctly specified for $\mathbb{E}[Y_i|D_i,X_i]$, or, $\hat{\pi}(D_i|X_i;\hat{\alpha})$ is correctly specified for $\pi(D_i|X_i)$. 

Thus, for treatment level $d \in \mathcal{D}$ we define $\hat{\mu}_{DR}(d)$ by
\begin{align}\label{mudr}
\hat{\mu}_{DR}(d)=\frac{1}{n}\sum_{i=1}^n\left[\Psi^{-1}\left\{m(d,X_i;\hat{\beta})\right\}+\frac{I_d(D_i)}{\hat{\pi}(d|X_i;\hat{\alpha})}\left[Y_i-\Psi^{-1}\left\{m(d,X_i;\hat{\beta})\right\}\right]\right],
\end{align}
Which we form by assuming a p.m.f. $f_{D|X}(d|x_i,\alpha)$, and estimating the parameter $\hat{\alpha}$ from a regression model using the observed treatment doses $D_i$ and covariates $X_i$. 
\end{Thm}

We can view the DR APO (\ref{mudr}) as a predicted estimating equation with residual bias correction \citep[e.g.][]{Kang/Schafer:2007}. Such an estimating equation can be formed by weighting or augmenting the regression model $\Psi^{-1}\left\{m(X_i,D_i;\beta)\right\}$ with $\widehat{\kappa}_i(d,X_i)=I_d(D_i)/\hat{\pi}(d|X_i;\hat{\alpha})$. Essentially, the argument is as follows. The augmented regression model wil be consistent if the OR model $\Psi^{-1} \left\{m(X_i,D_i; \beta)\right\}$ is correct for $\Expect(Y(d)|X)$ because the inclusion of the covariate $\widehat{\kappa}_i(D_i,X_i)$ simply adds noise to the predicted values, but leaves the consistency and asymptotic normality of the estimates unchanged. If the OR model is incorrectly specified, but the PS is correctly specified, the model will still be consistent because inclusion of the inverse PS gives rise to estimating equations that effectively correct for the bias in approximating $Y_i(d)$ using $\Psi^{-1} \left\{m(X_i,D_i; \beta)\right\}$ \citep[for details see][]{Bang/Robins:2005,Tsiatis:2006,Kang/Schafer:2007,Graham/et/al:2016}. For a proof of the DR property in the case of multivalued treatments see Appendix 1.   

The bias correcting estimating equations of the DR model can be derived via a number of standard regular asymptotically linear estimators. For instance, Maximum Likelihood Estimation (MLE), Maximum Quasi-Likelihood (MQL), Restricted MLE (REML) for linear mixed models (LMMs), and Penalised Quasi-Likelihood (PQL) for generalised linear mixed models (GLMMs) all provide estimating equations of the form
\begin{equation}
\sum_{i=1}^{n}\widehat{\kappa}_i(D_i,X_i)\frac{1}{\phi} \frac{\partial \left[ \Psi^{-1} \left\{m\left(D_i,X_i,\widehat{\kappa}_i(D_i,X_i); \xi\right)\right\} \right]}{\partial \xi^{\sf{T}}} \left[Y_i-\Psi^{-1} \left\{m\left(D_i,X_i,\widehat{\kappa}_i(D_i,X_i); \xi\right)\right\}\right]=0,
\end{equation}
where $\phi_i \equiv \phi (D_i,X_i)$ is a working conditional variance for $Y_i$ given $(D_i,X_i)$. Following the approach introduced by \citet{Scharfstein/et/al:1999}, a DR estimate of $\tau(d)$ can then be obtained as
\[\widehat{\tau}_{DR}(d)=\frac{1}{n}\sum_{i=1}^n\left[\Psi^{-1} \left\{m\left(d,X_i,\widehat{\kappa}_i(D_i,X_i); \hat{\xi} \right)\right\}-\Psi^{-1} \left\{m\left(0,X_i,\widehat{\kappa}_i(D_i,X_i); \hat{\xi}\right)\right\} \right].\] 

In this way, DR models add an additional layer of robustness in the sense that we only have to get one of the two component model right. Of course, it is always possible to get both the OR and PS models wrong, and moreover, and if the measure of $X$ available in the observed data is insufficient to guarantee the CIA, then both the OR and PS models will fail and DR estimation will not improve matters. For a Bayesian take on the DR approach see \citet{Graham/et/al:2016}.  

\subsection{Variance estimation for causal estimators}\label{VECE}

Finally in this section, we consider the issue of variance estimation. The estimators used to identify causal effects vary in nature and form. Some are based on particular parametric forms while others are based on non-parametric identification and mild functional form assumptions. In addition, as we have seen, many estimators involve multiple stages of calculation often with sequential plug-in of estimates. Under correct model specification, the APO / ATE estimates derived from these approaches are typically consistent and asymptotically normal (CAN) in the usual sense, e.g. 
\begin{align*}
\left(\widehat{\tau}_n-\tau \right)\overset{p}\longrightarrow 0\\
\sqrt{n}\left(\widehat{\tau}_n-\tau \right)\overset{d}\longrightarrow \mathcal{N} \left(0,\sigma^2_{\tau} \right),
\end{align*}
with zero asymptotic bias. It follows that standard errors and standard confidence intervals can be used for testing hypotheses; and that in general, variance estimation is justified by asymptotic theory, either using large sample approximations to the asymptotic variance via the delta method, or via bootstrap approximation. The Delta method, draws on Slutsky's theorem to provide a generalisation of the Central Limit Theorem (CLT) which allows us to derive approximate asymptotic results for functions of random variables. The bootstrap, developed by \citet{Efron:1979,Efron:1982}, is a technique for calculating the properties of estimators, such as the variance, by resampling the observed data with replacement to infer the population distribution. The standard asymptotic results these methods rely on are covered in detail in \citet{VanDerVaart:1998} and in popular textbooks on the theory of probability or statistical inference \citep[see for example][]{Casella/Berger:2002,Grimmett/Stirzaker:2001,Young/Smith:2005,Stuart/et/al:2007}.  

Considering first the Delta Method, and demonstrating in  the context of the OR ATE estimator, 
\begin{equation}\label{ORvar}
\hat{\tau}_{OR}=\frac{1}{N}\sum_{i=1}^N\left[m_1(d,x_i,\hat{\beta_1})-m_0(d,x_i,\hat{\beta_0}) \right],
\end{equation}
where $m_1(d,x_i,\beta_1)$ and $m_0(d,x_i,\beta_0)$ are parametric functions and $\hat{\beta_1}$ and $\hat{\beta_0}$ are $\sqrt{N}$ consistent and asymptotically normal estimators of the parameters of those functions; the asymptotic variance of $\hat{\tau}_{OR}$ can be expressed as 
\begin{align*}
\V \sqrt{N}(\hat{\tau}_{OR}-\tau_{ATE}) = &\Expect\left\{\left[m_1(d,x_i,\beta_1)-m_0(d,x_i,\beta_0) - \tau_{ATE}\right ]^2\right\}\\
& + \Expect \left[\nabla_{\beta_0}m_0(d,x_i,\beta_0) \right]V_{0}\Expect \left[\nabla_{\beta_0}m_0(d,x_i,\beta_0) \right]^\transpose\\
& +  \Expect \left[\nabla_{\beta_1}m_1(d,x_i,\beta_1) \right]V_{1}\Expect \left[\nabla_{\beta_1}m_1(d,x_i,\beta_1) \right]^\transpose
\end{align*}  
where $V_0$ and $V_1$ are the asymptotic variances of $\sqrt{N}(\hat{\beta}_0-\beta_0)$ and $\sqrt{N}(\hat{\beta}_1-\beta_1)$ respectively. The asymptotic variance of the OR ATE estimator can then be calculated using sample counterparts to the expectations shown in the expression above. 

Alternatively, applying the bootstrap, let $\hat{\tau}_{OR}(z)=\hat{\tau}_{OR}$ be the estimator obtained from data $z=(y,d,x)$, for which we want to know the variance. The nonparametric bootstrap proceeds as follows:
\begin{itemize}  
\item[(a)] Sample $z^*=(y^*,d^*,x^*)$ from $z$ and form the bootstrap replicate $\hat{\tau}_{OR}(z^*)$ using (\ref{ORvar}).
\item[(b)] Repeat (a) $B$ times to get $B$ independent bootstrap replicates $\hat{\tau}_{OR}(z^{*1})$,...,$\hat{\tau}_{OR}(z^{*B})$.
\item[(c)] Calculate the bootstrap variance using 
\[\V^*(\hat{\tau}_{OR})=\frac{1}{n^n - 1}\sum_{i=1}^{n^n}(\hat{\tau}_{OR_i}^*-\bar{\hat{\tau}}_{OR_i}^*)^2,\]
where $\hat{\tau}_{OR_i}^*$ is the estimator calculated from the $i$th resample and  $\bar{\hat{\tau}}_{OR_i}^*$ is the average of the estimator over all resamples. 
\end{itemize} 

A key advantage of the bootstrap for causal PS estimators is that the resampling scheme can address uncertainty and estimation error from multi-step estimation procedures by including all such steps within a single scheme. 

\section{Methods for estimation given a non-ignorable treatment assignment}\label{ENITA}
The validity of the estimation methods discussed in the previous section requires us to maintain that strong ignorability holds. In practice, this is often untenable, either because there are insufficient measured covariates to defend the CIA, or because other sources of endogeneity are at play, such as reverse causality or measurement error, inhibiting a causal interpretation of the data.

There are a number of popular estimators that are used under these conditions to obtain causal estimates of the APO and ATE.  Some use additional variables (instruments) to recover exogenous variation in treatments, while others exploit quasi-experimental conditions for identification. Here we review four of the most commonly used approaches: instrumental variables (IV), difference-in-differences (DID), synthetic control (SC) and regression discontinuity design (RDD).  

\subsection{Instrumental variables}   
Sometimes the presence of unmeasured confounding causes the CIA to fail with observed data. Under failure of the CIA it is not possible to achieve identification via OR, PS or DR based models. One solution that is commonly adopted in these circumstances, is to effectively bypass the CIA by introducing additional information in the form of instrumental variables (IVs), which allows us to abandon the requirement of conditional independence. 

The IV estimator is well known and widely used and for that reason we do not provide an extensive review here. We also focus on application of IV within the context of the linear model for a multivalued or continuous treatment variable ($\mathcal{D}\equiv(d_0,d_1,...,d_m)$ or $\mathcal{D} \subseteq  \mathbb{R}$), e.g. 
\begin{equation}\label{DGPLM1}
Y_i = \Expect[Y_i|D_i,X_i] + u_i = \tau D_i + \beta X_i + u_i
\end{equation}
since this is by far the most common setting for IV. An up--to-date discussions of IV for binary treatments, non-normal response distributions, and non-linear specifications is provided by \citet{Imbens:2014}.


Stacking observations for each individual we write the model for the DGP
\begin{align}\label{DGPLMa}    
Y|D,X & \sim \mathcal{N}(\tau D + \beta X, \sigma^2_Y)\\
D|X & \sim \mathcal{N}(\alpha X,\sigma^2_X) \nonumber 
\end{align}
We know from the discussion of OR based estimators above that if we have sufficient covariates $X$ to satisfy the CIA, then the OR model 
\begin{equation}\label{DGPLM2}    
Y=\tau D + \beta X + u.
\end{equation}
has a causal interpretation. When the CIA fails due to the presence of unmeasured confounders (e.g. we will assume $X$ is not observed), then this is no longer the case. For instance, say we estimate the model
\begin{equation}\label{OVBLM}
Y = D\tau  + e
\end{equation}
instead of the true model (\ref{DGPLM2}), then omission of the unobserved confounders, $X$, causes a problem of omitted variable bias (OVB) which produce bias and inconsistent estimates of the ATE. Such a failure amounts to violation of the Gauss-Markov condition that the error term is distributed independently of the regressors: $\Expect \left(D^{\sf{T}} e \right) \neq 0$. Sometime referred to as the population orthogonality condition, failure arises from the fact that $e= \beta X + u$, and by the definition of confounding, $\Expect \left(D^{\sf{T}} X \right) \neq 0$. 

The logic of IV is that since failure of the CIA implies that we have an identification problem, because $\Expect \left(D^{\sf{T}} e \right) \neq 0$, we will simply seek to directly remove this correlation from the model by introducing other observable variable(s) known as {\em instruments}, which we will denote $Z$. In this simple case with a single endogenous treatment variable, we will first assume that $Z$ is an $n \times 1$ vector. A valid instrument, e.g. one that will solve the identification problem, must satisfy two conditions 
\begin{enumerate}
{\bf \item Exclusion restriction} - the instrument must be uncorrelated with the error in the non-identified model: $\Expect \left(Z^{\sf{T}} e \right) = 0$. In other words, it must be exogenous, and would not feature in the non-identified model  
{\bf \item Relevance} - the instrument must be correlated with the the endogenous explanatory variable, conditionally on any other covariates: $\Expect \left(Z^{\sf{T}} D \right) \neq 0$ 
\end{enumerate}
The basic principle of IV is that we achieve identification by using the instruments to enforce orthogonality between the error term and an instrument transformed design matrix. The relationships assumed in IV estimation are shown graphically below.

\begin{figure}[h]
\centering \includegraphics[height=4.0cm]{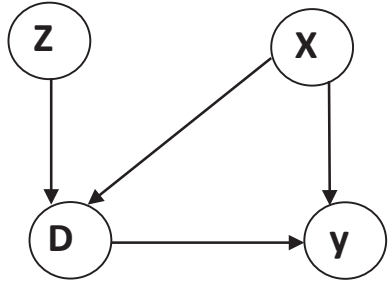}
{\caption{Relationships in Instrumental Variables estimation}}
\end{figure}

The defining characteristics of the IV model are that: changes in $Z$ are associated with changes in $D$, but do not lead to changes in $Y$ other than through $D$. $Z$ is causally associated with $D$ but {\it definitely} not with $Y$, and thus $Z$ would not be found in the regression model for $Y$.

To see how the IV conditions achieve identification, we premultiply our linear model $Y=D\tau+u$ by $Z^{\sf{T}}$  to get
\[Z^{\sf{T}}Y=Z^{\sf{T}} D \tau+Z^{\sf{T}}e.\]
Taking expectations and using the exclusion restriction $\Expect[Z^{\sf{T}}e]=0$, then 
\[\Expect[Z^{\sf{T}}Y]=\Expect[Z^{\sf{T}}D]\tau,\]
and we can solve for $\tau$ as
\[\tau_{IV}=\Expect[(Z^{\sf{T}}D)]^{-1}\Expect[(Z^{\sf{T}}Y)].\]
Thus, $\tau$ is identified using the IVs since the expectations $\Expect[(Z^{\sf{T}}D)]$ and $\Expect[(Z^{\sf{T}}Y)]$ can be consistently estimated using observed data. We can show this as follows. If $Z$ is uncorrelated with $e$ such that $\plim \ (n^{-1}Z^{\sf{T}}e)=0$, and, if $Z$ is associated with $D$ such that $\plim \ (n^{-1}Z^{\sf{T}}D)=\Sigma_{ZD}$ exists and is non-singular, then
\begin{align*}
Z^{\sf{T}}Y& =Z^{\sf{T}}D\tau+Z^{\sf{T}}e\\
n^{-1}Z^{\sf{T}}Y& =n^{-1}Z^{\sf{T}}D\tau+n^{-1}Z^{\sf{T}}e\\
\plim(n^{-1}Z^{\sf{T}}Y)&=\plim(n^{-1}Z^{\sf{T}}D)\tau+\plim(n^{-1}Z^{\sf{T}}e)
\end{align*}
Since $n^{-1}Z^{\sf{T}}e \overset{p}{\longrightarrow} \Expect[Z^{\sf{T}}e]= 0$ as $n \rightarrow \infty$, then using sample moments to consistently estimate $\Expect[(Z^{\sf{T}}Y)]$ and $\Expect[(Z^{\sf{T}}D)]$, we have
\[\hat{\tau}_{IV}=\left(n^{-1} Z^{\sf{T}}D\right)^{-1}n^{-1}Z^{\sf{T}}Y=\frac{\Cov(Z,Y)}{\Cov(Z,D)}.\]

So far we have worked with a single endogenous treatment variable and a single instrument, but the IV set up generalises immediately to one in which $\dim Z = \dim D \geq 1 = M$. More generally, the so called order condition for IV requires that the number of instruments be greater than or equal to the number of endogenous covariates: e.g. $\dim Z = L \geq \dim D = M$. In fact, exogenous covariates can act as instruments in their own right since they satisfy the two IV validity conditions. So we simply need to find as many (or more) valid instruments are there are endogenous covariates. When $L=M$ the model is said to be just-identified. When $L > M$ the model is said to be overidentified. 

The IV estimator described above requires that $L=M$. In the overidentified setting we could simply discard some instruments, but this can result in loss of efficiency. Instead, both just identified and over-identified model are usually estimated via two-stage Least Squares (2SLS), which proceeds as follows
\begin{enumerate}
\item[1.] Regress each column of $D$ on the instrument matrix $Z$ and save the predicted values $\hat{D}=Z(Z^{\sf{T}}Z)^{-1}Z^{\sf{T}}D=P_Z D$.
\item[2.] Regress $Y$ on the predicted values from the first stage: $Y=\hat{D}\tau + e$.
\end{enumerate}
The resulting 2SLS estimator is
\begin{align*} 
\tau_{2SLS}& =[(D^{\sf{T}}Z(Z^{\sf{T}}Z)^{-1}Z^{\sf{T}}D)]^{-1}[(D^{\sf{T}}Z(Z^{\sf{T}}Z)^{-1}Z^{\sf{T}}Y)]\\
& \tau + [(D^{\sf{T}}Z(Z^{\sf{T}}Z)^{-1}Z^{\sf{T}}D)]^{-1}[(D^{\sf{T}}Z(Z^{\sf{T}}Z)^{-1}Z^{\sf{T}}e)],
\end{align*} 
where the second term disappears in expectation due to $\Expect[Z^{\sf{T}}e]= 0$. Asymptotic results can be derived by applying a LLN to the sample moments.  

IV can be used to establish causal effects under a non-ignorable treatment assignment and is particularly useful when endogeneity via bi-directionality is present.  However, it is crucial that the two key assumptions of exogeneity and relevance are met, and in practice valid instruments are hard to find. When instruments are only weakly correlated with the endogenous regressors, or when the instruments themselves are correlated with the error term, IV estimation can produce severely biased and inconsistent estimates. This problem is further confounded by the fact that the available diagnostic statistics do not provide a full proof means for detecting an inadequate instrument specification. To quote \citet{Hahn/Hausman:2003}, even using standard tests for instrument validity ``the researcher may estimate `bad results' and not be aware of the outcome'' (p 118). 

When panel data are available, but exogenous instruments are not, one commonly adopted route forward is to use the dynamic panel Generalised Method of Moments (GMM) estimator for panel data. Dynamic GMM applies differencing to remove the unobserved individual effects and uses the time series nature of the data to derive a set of instruments from lagged levels which are assumed correlated with the differenced covariates but orthogonal to the errors. A set of moment conditions can then be defined and solved within a GMM framework which will be satisfied at the true value of the parameters to be estimated. 

\subsection{Difference-in-differences}  

Differences-in-differences (DID) is a `before and after' treatment effect estimation approach that is applicable when the effect of treatment on units can be represented as a binary variable. It can reveal impacts associated with exposure to an intervention relative to non-exposure (control), but it cannot tell us about impacts by scale or `dose' of intervention. For DID we therefore maintain that $\mathcal{D}\in\{0,1\}$. 

Say we want to estimate an ATE in a before-after setting. An identification problem will arise if unobserved confounding is present in the form of differences between the treated and untreated units which affect outcomes and are also influential in treatment assignment. If so, ignorability fails and by extension identification via OR, PS or DR models. In addition to unobserved confounding, there may be temporal trends that affect the outcome variable due to events that are unrelated to the treatment. 

The DID estimator addresses such potential sources of bias by using information for both treated and control groups in both pre and post treatment periods. In the basic DID model, we model the outcomes, $Y_{it}$, for units $i$, $i=(1,2,...,N)$ in binary time periods $t \in \{0,1\}$ (with $t=0$ representing the pre-treatment period and $t=1$ the post-treatment period) using
\begin{equation}
Y_{it}= \mu+\alpha I_{1}(D_i)+\delta_t \cdot t+\tau_{D} \cdot I_{1}(D_i)\cdot t+\varepsilon_{it},
\label{eq18}
\end{equation}  
where $\upsilon_{it}$ is a potentially autoregressive error with mean zero in each time period. The effect of the treatment is captured by the parameter $\tau_D$, which provides the sample counterpart to 
\begin{equation}
\tau_{D}=\Expect[Y_i(1)] - \Expect[Y_i(0)]=\left\{ \Expect [Y_{i,1}|I_{1}(D_i)]- \Expect [Y_{i,0}|I_{1}(D_i)]\right\} - \left\{\Expect [Y_{i,1}|I_{D_{i}}(0)]- \Expect [Y_{i,0}|I_{D_{i}}(0)]\right\},
\label{eq20}
\end{equation}  
with least squares estimate
\[ \hat{\tau}_{D}= \left(\overline{Y}_{11}-\overline{Y}_{10}\right)-\left(\overline{Y}_{01}-\overline{Y}_{00}\right),\]
where $\overline{Y}_{11}$ is the sample average outcome for treated units in year 1. 

The `double-differencing' of the DID estimator removes two potential sources of bias. First, it eliminates biases in second period comparisons between the treated and controlled groups that could arise from time invariant characteristics. Second, it corrects for time varying biases in comparisons over time for the treated group that could be attributable to time trends unrelated to the treatment.

There are two key identifying assumptions required of the the basic DID model. First, we assume that the treatment assignment and the error $\varepsilon_{it}$ are independent
\[\P(I_{1}(D_i)|\varepsilon_{it})=\P(I_{1}(D_i)),\]
for $t=0,1$. 

Second, we assume that the average outcomes for the treated and control groups would have followed parallel paths over time in the absence of the treatment. If $Y_{it}(0)$ is the outcome that unit $i$ experiences in time $t$ in the absence of treatment, then for binary period $t\in{0,1}$ we make the following assumption. 
\begin{Assum} (Unconditional parallel outcomes). For identification of treatment effects in the basic DID model it is necessary that the average outcomes for the treated and control groups would have followed parallel paths over time in the absence of the treatment
\[
\Expect_i\left[Y_{i,1}(0)-Y_{i,0}(0)|I_{1}(D_i)]=\Expect[Y_{i,1}(0)-Y_{i,0}(0)|I_{0}(D_{i})\right].
\]
\label{DID}
\end{Assum}

There are a number of useful extensions of the DID model that can help render the assumptions required for identification more plausible.
\begin{enumerate}
{\bf \item DID with covariates - } Covariates ($X_i$) can be added to the DID model to adjust for heterogeneity in outcome dynamics between treated and control groups. With the vector of covariates included the DID model is
\begin{equation}
Y_{it}= \mu+X_{i}^{'}\beta+\alpha I_{1}(D_i)+\delta_t \cdot t+\tau_{D} \cdot I_{1}(D_i)\cdot t+\varepsilon_{it}.
\label{eq21}
\end{equation}    
The impact of the treatment across the population can be generalised by allowing for interactions between $X_i$ and $I_{1}(D_i)$.

The addition of covariates that are thought to be associated with the dynamics of the outcome variables can be particularly useful in satisfying the identifying restrictions of the DID model. Thus, regarding independence of the error, we can condition on covariates predetermined at $t=0$ such that 
\[\P(I_{1}(D_i)|X_i,\varepsilon_{it})=\P(I_{1}(D_i)),\] 
is required rather than unconditional independence. Similarly, in the covariate model assumption (\ref{DID}) becomes
\[\Expect_i[Y_{i,1}(0)-Y_{i,0}(0)|X_i,I_{1}(D_i)]=\Expect[Y_{i,1}(0)-Y_{i,0}(0)|X_i,I_{0}(D_{i})],\] 
which may be considerably more plausible. 

{\bf \item DID with multiple groups and time periods - } When the number of time periods exceeds 2 the DID model can be reformulated by including a vector of time dummies and an indicator variable which takes a value of 1 for groups and time periods that were subject to the treatment
\begin{equation}
Y_{it}= \mu+X_{it}^{'}\beta+\alpha I_{1}(D_i)+\sum_{t=1}^T\delta_t \cdot I_{T_{i}}(t) +\tau_{D} \cdot I_{i}+\varepsilon_{it}.
\label{eq22}
\end{equation}    
where $I_{T_{i}}(t)$ is an indicator variable for year $t$ and $I_{i}$ is the group-time treatment indicator. 

For multiple time periods and groups the model is
\begin{equation}
Y_{it}= \mu+X_{it}^{'}\beta+\sum_{g=1}^G\alpha_g I_{G_{i}}(g)+\sum_{t=1}^T\delta_t \cdot I_{T_{i}}(t) +\tau_{D} \cdot I_{i}+\varepsilon_{it},
\label{eq23}
\end{equation}    
where $I_{G_{i}}(g)$ is an indicator variable for year membership of group $g$.

{\bf \item Longitudinal DID - } So far DID has been discussed in a repeated measures setting in which we have a random sample of observations in each time period. If instead the data are for the same units over time the  other formulations are possible. For instance, the basic DID model can then be estimated in differences as
\[
Y_{i,1}-Y_{i,0}= \delta +\tau_D I_{1}(D_i)+(\varepsilon_{i,1}-\varepsilon_{i,0}).
\]
Furthermore. if we can assume unconfoundedness given lagged outcomes, then a differenced estimator could be $Y_{i,1}-Y_{i,0}= \delta +\tau_{DU} I_{1}(D_i)+ Y_{i,0}+\varepsilon_{it}$. To estimate this model we must either be able to assume that $\Expect[\varepsilon_{it}|Y_{i,0}]=0$ or be able to correct for any correlation through, for instance, the use of lagged differences in the outcome variable as instruments in a Generalized Method of Moments estimator. 
\end{enumerate} 

It is important to note two potential limitations with the DID approach. First, it relies on the strong identifying assumption that the average outcomes for the treated and control groups would have followed parallel paths over time in the absence of the treatment. Second, the model is sensitive to error specification, and in particular, it has been shown that the existence of correlation within groups or over time periods can adversely affect the performance of the DID estimator \citep[e.g.][]{Bertrand/et/al:2004}.

\subsection{Synthetic control}  

The synthetic control (SC) approach can be thought of a development of DID which aims to improve comparability of the treatment and control groups \citep[e.g.][]{Abadie/Gardeazabal:2003,Abadie/et/al:2010,Abadie:2021}. This is done by imposing weights for multiple control groups that render then more similar to the treated groups. The weights can be chosen by considering the distances in averages between treated and untreated groups or by using information on group level covariates.

Let $J$ be the number of available control groups and let $K$ be the number of characteristics we observe for each group. Our objective is to assign an appropriate weight to each control, such that the weighted SC will better approximate the observed characteristics of the treated group. The optimal $J \times 1$ vector of weights $W=(w_{1},.., w_{J})$ is obtained from the following minimization problem
\begin{equation}
\min_{W} (X_{1}-X_{0}W)^{\sf{T}}V(X_{1}-X_{0}W)
\end{equation}
such that $ \sum_{i=1}^{J} w_{i} = 1$, and where $X_{1}$ is a $K \times 1$ matrix containing all the characteristics of the treated group and $X_{0}$ is the $K \times J$ vector containing the characteristics of the control groups. 

The solution to this problem is a vector of optimal weights, $W^{*}(V)$, which add up to one. The optimal vector depends on $V$, a $K \times K$ diagonal matrix that assigns weights to each characteristic. We choose the matrix $V$ so that the outcome of interest in the treated group is best replicated by its synthetic counterpart in the period prior to treatment. Let $T_{p}$ be the number of years observed prior to the treatment, then the problem of choosing V can be specified as follows 
\begin{equation}
\min_{V} (Z_{1}-Z_{0}W^{*}(V))^{\sf{T}}(Z_{1}-Z_{0}W^{*}(V))
\end{equation}
where $Z_{1}$ is a $T_{p} \times 1$ vector with a time series of the variable of interest in the treated group, and $Z_{0}$ is a $T_{p}\times J$ vector with a time series of the variable of interest in the control groups.

Once we have derived the SC group, we compare its evolution with respect to the treated group during the years following the treatment. Let $T$ be the number of years observed after the treatment, then we are interested in the gap between the treated group and its synthetic analogue
\begin{equation}
 Y_{1}-Y_{0}W
\end{equation}
where $Y_{1}$ is a $T \times 1$ vector with a time series of the variable of interest in the treated group for the post treatment years, and $Y_{0}$ is a $T \times J$ vector with a time series of the variable of interest in the control groups for the post treatment years. 

In this way, the SC approach allow us to identify not only a single ATE estimate, but also reveals how the ATE effect evolves over time post-treatment. Clearly, the accuracy of the resulting estimator depends crucially on the choice of weights, and thus on the set of characteristics used to form the weights in the first place. \citet{Abadie:2021} and \citet{Athey/Imbens:2017} provide a detailed discussion of these issues and how improvements can potentially be made in particular settings. 

\subsection{Regression discontinuity designs}

RDD is a useful approach under non-ignorability when a given covariate, referred to as the forcing or running variable, partly or completely determines treatment assignment \citep[for reviews of RDD see][]{Imbens/Lemieux:2008,Lee/Lemieux:2010}. Under a `sharp' RDD design the conditional probability of assignment to treatment is of size one at a chosen threshold of the forcing variable, while under a `fuzzy' design the probability at the threshold is less than one. The RDD method works by exploiting discontinuity in treatment assignment to estimate the conditional distributions of outcome either side of the threshold. A discontinuity in these conditional distributions is interpreted as evidence of a causal treatment effect.

We will consider identification under a sharp RDD design in which treatment assignment is a deterministic function of the forcing variable, which we denote by $T$. The treatment status of unit $i$ is given by
\[D_{i} = 1[T_i \geq c]\]
where $1[T_i \geq c]$ is an indicator function that is equal to one if the statement in brackets is true or zero otherwise. Our aim is to estimate
\[\tau_{SRD}=\Expect \left[Y_{i}(1) - Y_{i}(0)|T_i = c \right]=\Expect\left[Y_{i}(1)|T_i = c \right]-\Expect\left[Y_{i}(0)|T_i = c \right],\]
which is equivalent to the population ATE if the treatment effect is constant. Due to unobserved potential outcomes we cannot observe both expectations. Instead, we assume continuity of the expectations in $T$ 
\[\Expect\left[Y_{i}(0)|T_i = c \right]=\lim_{t \uparrow c}\Expect\left[Y_{i}(0)|T_i = t \right]=\lim_{t \uparrow c}\Expect\left[Y_{i}|T_i = t \right],\] 
which implyies that
\[\tau_{SRD}=\lim_{t \downarrow c}\Expect\left[Y_{i}|T_i = t \right]-\lim_{t \uparrow c}\Expect\left[Y_{i}|T_i = t \right].\]

The estimated ATE under sharp RDD is simply the difference in the conditional expectation of the outcome either side of the discontinuity. For this calculation to produce valid inference, the key identifying assumption is there is a discontinuous change in the probability at the threshold. 

A valid RDD design involves fairly minimal assumptions. It will provide consistent causal estimates of the ATE without the need to condition on baseline covariates, but it can be useful to include them to reduce the sampling variability of the estimator and improve precision. The running variable can be any observed covariate, and actually in some applications is simply time \citep[e.g.][]{Singh/Graham:2022,Ma/et/al:2021}. 

\section{Simulations}\label{SIMS}

In this section we demonstrate application of the causal methods reviewed above in engineering research. The examples are drawn from my own field of transportation engineering, but have have general applicability across DCE applications \citep[for a discussion of causal inference in the specific context of ex-post transport program evaluation see][]{Graham:2014}. We use Monte Carlo simulation to represent DGPs typical of those we encounter in real world data. Each simulation is based on 1,000 runs on generated datasets of size 1,000. In each case, we report mean values and variances of the estimates obtained and the mean squared error (MSE). We use standard distributions to generate the relationship described in the DGPs, e.g. Normal ($\mathcal{N}$), Bernoulli ($\mathcal{B}$) and Uniform Continuous ($\mathcal{U}$). We will not worry about the units of the generated variables or introduce truncation to impose realism, since the models could be specified with variables in logs or transformed in some other way, so actual variable values make no difference. {\tt R} code for the simulations is available for readers to run and adapt, and citations are provided to real world research.   

\subsection{Case study 1: treatment effect estimation under strong ignorability}

We are interested in the effect that speed cameras (SCs) have on road traffic collision rates (RTC) \citep[e.g.][]{Graham/et/al:2019,Li/Graham:2016,Li/et/al:2013}. We have data at the link level on both of these variables for treated and control links. We believe that the assignment of speed cameras is confounded, since they are allocated according to a set of criteria that are also relevant for RTCs.

Our response of interest $Y \subseteq  \mathbb{R}$ is RTC, our binary treatment $D \in\{0,1\}$ measure the presence of an SC, and our covariate $X \subseteq  \mathbb{R}$ represents confounders. The DGP is given by
\begin{align*}
X & \sim \mathcal{N}(0,10)\\
D|X & \sim   \mathcal{B}(\text{expit}(\alpha_0+\alpha_1 X))\\
Y|D,X & \sim  \mathcal{N}(\beta_0 + \tau D + \beta_1 X, 5)
\end{align*}
where $\alpha_0=2$, $\alpha_1=0.5$, $\beta_0=10$, $\tau=-5$, $\beta_1=0.5$. The true ATE $\Expect\left[Y_i(1)-Y_i(0)\right]$ is given by parameter $\tau = -5.0$.

The following causal estimators are simulated
\begin{itemize}
\item[1.] $\widehat{\tau}_{OR1}$ - an OR model based on the correctly specified model: $\Expect[Y|D,X]=\beta_0+\tau D+\beta_{1}X$. 
\item[2.] $\widehat{\tau}_{OR2}$ - same as [1.] except based based on an incorrectly specified OR model with covariate $X$ excluded.
\item[3.] $\widehat{\tau}_{PS1}$ - an IPW PS estimator based on a correctly specified PS model estimated using $\Expect[D|X]=\alpha_0+\alpha_{1}X$.  
\item[4.] $\widehat{\tau}_{PS2}$ - an IPW PS estimator based on an incorrectly specified PS model, which is generated randomly from the normal distribution $\widehat{\pi}(D|X) \sim \mathcal{N}(\bar{\pi},0.5)$ with mean equal to the mean of the true PS ($0.01-0.99,$) and truncated between 0.01 and 0.99.
\item[5.] $\widehat{\tau}_{DR1}$ - a DR model based on an incorrectly specified OR model ($X$ excluded), but with PS weighting based on the correct PS model
\item[6.] $\widehat{\tau}_{DR2}$ - a DR model based on a correctly specified OR model, but with weighting based on an incorrect PS model (see 3. above).
\item[7.] $\widehat{\tau}_{DR3}$ - a DR model based on an incorrectly specified OR model weighted with weights based on an incorrect PS model.
\end{itemize}

Table \ref{sims1} shows our simulation results (see also Figure \ref{CSSimF} in the appendix). 
 
\begin{table}[htbp]
  \centering
  \caption{Simulation of causal ATE estimators under strong ignorability ($\tau=-5.0$)}
    \begin{tabular}{lccc}
    \hline
          & Av. Est. & Emp. Var. & MSE \\
    \hline
    OR1   & -4.999 & 0.084 & 0.084 \\
    OR2   & -3.149 & 0.087 & 3.512 \\
    PS1   & -4.949 & 2.054 & 2.054 \\
    PS2   & 3.697 & 2.859 & 78.499 \\
    DR1   & -4.944 & 0.398 & 0.400 \\
    DR2   & -4.977 & 0.309 & 0.309 \\
    DR3   & -3.125 & 0.313 & 3.827 \\
    \hline
    \end{tabular}%
  \label{sims1}%
\end{table}%

The correctly specified OR model, $\widehat{\tau}_{OR1}$, provides a good approximation to the true value of $\tau$. The incorrectly specified OR model, OR2, fails to satisfy the CIA and consequently $\widehat{\tau}_{OR2}$ provides a poor approximation to the true ATE. A good estimate of $\tau$ is achieved via the correctly specified PS model ($\widehat{\tau}_{PS1}$), but when the PS is model is mispecified ($\widehat{\tau}_{PS2}$) the estimate of the ATE is far away from the true value. This tendency of the inverse PS model to fail quite considerably under severe misspecification is well known in the literature \citep[e.g.][]{Kang/Schafer:2007}.  Weighting the incorrectly specified OR model with weights $\widehat{\kappa}(D,X)$, based on a correctly specified PS model, as in the DR1 model, provides correction for misspecification bias with an average point estimate very close to the true value. The DR2 model also produces valid point estimates because weighting by weights based on an incorrectly specified PS model does not does not induce bias when the OR model is correct, but it does increase variance. Finally, if both the OR and PS models are wrongly specified, as in DR3, the model fails to produce a consistent estimate of the mean ATE.  

\subsection{Treatment effect estimation with unobserved confounding and panel data}

\subsubsection{Case study 2: Panel data simulation with time-invariant confounding}

In a longitudinal data setting we are interested in the effect that road capacity expansions $D_{it}$ have on network speeds $Y_{it}$ \citep[e.g.][]{Graham/et/al:2014}. We have disaggregate data on commute distance and on productivity, but there are a number of time-invariant worker level characteristics, $W_i$, that we do not observe which simultaneously affect treatment and response. The DGP for this case study is   
\begin{align*}
W_i & \sim \mathcal{U}(1,100)\\
D_{it} & = \delta W_i+\mathcal{N}(\mu_{D}, \sigma^2_{D})\\
Y_{it} &  = \alpha + \tau D_{it} + \gamma W_{i}+\mathcal{N}(0, \sigma^2_{e})
\end{align*}
where $\delta=2$, $\alpha=1.0$, $\tau=0$, $\gamma=2$. The true ATE $\Expect\left[Y_{it}(1)-Y_{it}(0)\right]$ is given by parameter $\tau = 0$, and there is therefore no causal effect.

Five models are simulated: POLS, RE, FD, FE and CRE. The results are shown in table \ref{PD1Sim} below (see also Figure \ref{PD1SimF} in the appendix). 
\begin{table}[htbp]
  \centering
  \caption{Simulation of panel ATE estimators with time invariant confounding ($\tau= 0$)}
    \begin{tabular}{lccc}
    \hline
          & Av. Est. & Emp. Var.  & MSE \\
    \hline
 	POLS  & -0.993 & 0.000 & 0.985 \\
 	RE    & -0.991 & 0.000 & 0.983 \\
    FD   & -0.003 & 0.010 & 0.010 \\
    FE & -0.001 & 0.004 & 0.004 \\
    CRE & -0.0001 & 0.004 & 0.004 \\
    \hline
    \end{tabular}%
  \label{PD1Sim}%
\end{table}%

We do not observe $W_i$, and since it is a confounder, its omission will cause bias in estimate of the ATE unless we use appropriate panel adjustment. This effect is demonstrate in the results. We find substantial estimation bias in the POLS and RE models which assume $\Cov(D_{it},W_i)=0$. These model estimate a significant reduction in speeds from road capacity expansions, when there is in truth no effect. The FD, FE and CRE models, on the other hand, are able to consistently estimate the ATE, net of time-invariant confounding, which is swept out of the model through differencing or demeaning. Note that the CRE and FE models produce identical results, and this because the two approaches actually lead to the same estimator for the ATE in this simple example \citep[][]{Wooldridge:2019}.  

\subsubsection{Case study 3: Panel data simulation with time-varying confounding}

We now consider the same panel data setting, but modify the DGP to one one in which confounding varies over time.
\begin{align*}
W_{i} & \sim \mathcal{U}(1,100)\\
W_{it} & = W_i+ \mathcal{N}(\mu_{W},\sigma^2_{W})\\
D_{it} & = \delta W_{it}+\mathcal{N}(\mu_{D}, \sigma^2_{D})\\
Y_{it} & =\alpha + \tau D_{it} + \gamma W_{it}+\mathcal{N}(0, \sigma^2_{e})
\end{align*}

The results are shown in table \ref{PD2Sim} below (see also Figure \ref{PD2SimF} in the appendix). 

\begin{table}[htbp]
  \centering
  \caption{Simulation of panel ATE estimators with time varying confounding ($\tau=-2.0$)}
    \begin{tabular}{lccc}
    \hline
          & Av. Est. & Emp. Var.  & MSE \\
    \hline
    POLS  & -0.990 & 0.000 & 0.980 \\
    RE    & -0.989 & 0.000 & 0.988 \\
    FD   & -0.197 & 0.008 & 0.046 \\
    FE & -0.394 & 0.002 & 0.157 \\
    CRE & -0.394 & 0.002 & 0.157 \\
    \hline
    \end{tabular}%
  \label{PD2Sim}%
\end{table}%

A priori we know that if $W_{it}$ is omitted from the model for $Y_{it}$, bias in estimation of $\tau$ results. This is demonstrated in the simulation results: all of the panel estimators produce biased estimates of the ATE. The panel models that adjust for time-invariant confounding tend to do better, because they at least control for bias related to omitted cross sectional heterogeneity, but they still fail to correct adequately for time varying bias because the CIA fails.     

\subsection{Treatment effect estimation under non-ignorable treatment assignment}

\subsubsection{Case study 4: Instrumental variables estimation with a continuous treatment}

In a cross-sectional study, we aim to estimate the effect that urban density, $D_i$, has on automobile use, $Y_i$ \citep[][]{Vance/Hedel:2007}. There are a set of unobserved confounders, represented by $X_i$ that simultaneously determine urban form and automobile use and their exclusion from the model will induce OVB. However, we have available an instrument, $Z_i$, which measures underlying geological characteristics. The DGP for this case study is given by
\begin{align*}
X_i & \sim \mathcal{N}(15,1)\\
Z_i & \sim \mathcal{N}(0,1)\\
D_i|X_i,Z_i & \sim \mathcal{N}(\alpha_0+\alpha_1 X_i+\alpha_1 Z_i, \sigma^2_{D_i})\\
Y_i|D_i,X_i & \sim  \mathcal{N}(\beta_0 + \tau D_i + \beta_1 X_i, \sigma^2_{Y_i})
\end{align*}
where $\alpha_0=1$, $\alpha_1=0.5$, $\alpha_2=1$, $\beta_0=1$, $\tau=-1$, $\beta_1=0.5$. The true ATE $\Expect\left[Y_i(1)-Y_i(0)\right]$ is given by parameter $\tau = -1.0$.

We simulate 4 models
\begin{itemize}
\item[1.] $\widehat{\tau}_{OR_t}$ - based on the correctly specified OR model: $\Expect[Y|D,X]=\beta_0 + \tau D + \beta_1 X$. 
\item[2.] $\widehat{\tau}_{OR_n}$ - based on an incorrectly specified OR model with covariate $X$ excluded.
\item[3.] $\widehat{\tau}_{IV1}$ - IV estimation via 2SLS with incorrect model $\Expect[Y|D,X]=\beta_0 + \tau D $ and valid instrument $Z$ (correlated with urban form but not with automobile use, and therefore relevant and exogenous).   
\item[4.] $\widehat{\tau}_{IV1}$ - same as 3. but using an invalid instrument that is correlated with $D$ (e.g. non-exogenous).  
\end{itemize}

The results are shown in table \ref{IVSimT} below (see also Figure \ref{IVSimF} in the appendix).  

\begin{table}[htbp]
  \centering
  \caption{Simulation of IV estimator under non-ignorability ($\tau=-1.0$)}
    \begin{tabular}{lccc}
    \hline
          & Av. Est. & Emp. Var.  & MSE \\
    \hline
    OR (correct) & -1.000 & 0.001 & 0.001 \\
    OR (naïve) & -0.800 & 0.001 & 0.041 \\
    IV1    & -1.000 & 0.001 & 0.001 \\
    IV2 &-0.501 & 0.002 & 0.252 \\
    \hline
    \end{tabular}%
  \label{IVSimT}%
\end{table}%

The correct OR model, which includes $X$ in the regression, attains an unbiased estimate of the ATE. The naïve model excludes $X$ and makes no further adjustment, leading to a poor estimate of $\tau$. The IV1 model, estimated via 2SLS using a valid instrument, nullifies the effect of confounding bias and consequently produces a consistent ATE estimate. Use of an endogenous instrument, as in IV2, produces an ATE estimate with larger bias than the naïve OLS model.            

\subsubsection{Case study 5: Difference-in-Differences estimation with a binary treatment}

We are interested in the effect that the introduction of congestion charging has on traffic flows \citep[e.g.][]{Ouali/et/al:2021}. We have data before and after the congestion charge was introduced for zones that were subject to the charge and for those that were not. To estimate the causal effect we use a difference-in-difference model. In a two period setting, $t=(0,1)$, the DGP is given by
\begin{align*}
X_0 & \sim \mathcal{N}(0,1)\\
D_1|X_0 & \sim   \mathcal{B}(\text{expit}(\alpha X_0))\\
Y_0|X_0,D_1 & \sim \mathcal{N}(\beta_{D_0} + D_1 + \beta_{X_0} X_0, \sigma^2_{Y_0})\\ 
Y_1|X_0,D_1 & \sim \mathcal{N}(\beta_{D_1} + D_1 + \tau D_1 + \beta_{X_0} X_0, \sigma^2_{Y_1})\\     
\left(Y_1-Y_0\right)|D_1 = \Delta Y_1|D_1 & \sim  \mathcal{N}\left(\left(\beta_{D_1}-\beta_{D_0} \right) + \tau D_1, \sigma_{\Delta Y}\right)
\end{align*}
where $\tau=-4$ and $\beta_{D_1}-\beta_{D_0}=2$. Note that the time invariant confounders $X_0$ are removed via differencing and the parallel trend is satisfied. 

We simulate two DID models
\begin{enumerate}
\item DID1 - based on a correctly specified DID regression model: 
\[\Expect[Y|D]=\gamma_0+\gamma_1 D_1 + \gamma_2 T+ \tau_D (D_1 \times T)+\epsilon\]
\item DID2 - a modified DGP in which the parallel trend is violated by introducing a covariate that influences only control in the post-treatment period, e.g. 
\[Y_1|X_0,X_1,D_1 \sim \mathcal{N}(\beta_{D_1} + D_1 + \tau D_1 + \beta_{X_0} X_0+ \beta_{X_1} X_1 \times (1-D_1), \sigma^2_{Y_1}).\] 
Again the DID regression we use is:
$\Expect[Y|D]=\gamma_0+\gamma_1 D_1 + \gamma_2 T+ \tau_D (D_1 \times T)+\epsilon$   
\end{enumerate}
 
The results are shown in table \ref{DIDSim} below (see also Figure \ref{DIDSimF} in the appendix). 
\begin{table}[htbp]
  \centering
  \caption{Simulation of DID estimator ($\tau=-4.0$)}
    \begin{tabular}{lccc}
    \hline
          & Av. Est. & Emp. Var.  & MSE \\
    \hline
    DID1  & -4.005 & 0.008 & 0.008 \\
    DID2  & -5.001 & 0.010 & 1.003 \\
    \hline
    \end{tabular}%
  \label{DIDSim}%
\end{table}%

The DID model produces a consistent estimate of $\tau$ when the parallel trend holds. Note, however, that the simulation demonstrates the necessity of this assumption as shown in the bias of DID2, e.g. $\V(\tau) \neq MSE(\tau)$ .  

\subsubsection{Case study 6: Regression Discontinuity Design estimation with a binary treatment}

We are interested in evaluating the effect of imposition of a low emission zone (LEZ) on air quality \citep[e.g.][]{Ma/et/al:2021}. We have data before and after the imposition of the LEZ, and we use a sharp RDD model to evaluate its effect. The DGP for this case study is given by
\begin{align*}
T & \sim \mathcal{U}(-1, 1)\\
Y & \sim \mathcal{N}\left(\alpha + \beta T + \tau \times (T \geq 0), \sigma^2_{Y} \right)
\end{align*}
where assignment to treatment, e.g. $D=1[T \geq c]$, is determined by some threshold $c=0$ of the forcing variable, which in this case could be time. The true ATE is given by parameter $\tau = 5.0$.

We simulate three RDD models
\begin{enumerate}
\item RDD1 - based on a correctly specified sharp RDD regression model:
\[Y=\alpha+\tau D+\beta_{1}(T-c)+\beta_{2}D(T-c)+\epsilon\]
\item RDD2 - a modified DGP in which fuzziness is added to the threshold to simulate the presence of other factors that determine assignment to the treatment. Again, we estimate using a sharp RDD regression design, effectively ignoring the fact that treatment status is only partially determined by the running variable.
\item RDD3 - same DGP as RDD2, but this time using a fuzzy RDD regression.       
\end{enumerate} 

The results are shown in table \ref{RDDSim} below (see also Figure \ref{RDDSimF} in the appendix). 
\begin{table}[htbp]
  \centering
  \caption{Simulation of RDD estimator under non-ignorability ($\tau=5.0$)}
    \begin{tabular}{lccc}
    \hline
          & est tau & var tau & MSE tau \\
    \hline
    RDD1  & 4.995 & 0.015 & 0.015 \\
    RDD2  & 3.996 & 0.015 & 1.023 \\
    RDD3  & 4.995 & 0.024 & 0.024 \\
    \hline
    \end{tabular}%
  \label{RDDSim}%
\end{table}%

Estimates of $\tau$ are unbiased under the sharp RDD estimator. When the DGP is modified to invalidate the assumption of continuity at the threshold, the sharp RDD model fails to consistently estimate the true ATE. However, this can be rectified, as in RDD3, by switching to a fuzzy RDD model.  

\section{Conclusions}\label{CONCS}

In this paper we have reviewed methods that seek to draw causal inference from observational data, and have demonstrated how they can be applied to empirical problems in engineering research. Causal inference aims to quantify effects that occur due to explicit intervention (or `treatment') in non-experimental settings, often for non-randomly assigned treatments. Such conditions are frequently encountered in engineering research, and for that reason, causal inference has immediate and valuable applicability in our field. 

The paper has outlined a framework for causal inference based on the concept of potential outcomes, and has discussed methods that can be used for estimation. While these can be used to successfully derive inference about the causal effects of engineering interventions, there are several practical challenges that must be acknowledged.  

First, is that for some methods (reviewed in section 4), all confounders must be observed and measured to obtain valid causal inference. This stringent data requirement must be met to satisfy the conditional independence assumption (CIA). In practice, we often work in settings with unobserved confounders, or with measurement error, leading to potential failure of the CIA. Furthermore, the problem of unobserved confounding is exacerbated by the fact that there are no reliable diagnostic procedures to comprehensively test for conditional independence. 

Second, there are other methods (reviewed in section 5), which although perhaps less onerous in terms of data requirements, instead impose other stringent identifying assumptions. For IV estimation our instruments must be valid (e.g. relevant and exogenous), for DID models we must be able to defend a parallel trend, while continuity of outcomes and non-manipulation of the treatment threshold must hold for valid identification under RDD. It can be hard to establish empirically whether these assumptions hold in practice, and again diagnostics are not always clear cut.

A third major challenge for causal inference in engineering applications relates to the SUTVA, which must be met for all methods reviewed in the paper. A key implication of the SUTVA is that for each unit, outcomes must be independent of the treatment status of other units, or in other words, there must be no `interference' in treatment between units. The assumption of no interference is often satisfied naturally when units are physically separated or otherwise have no means of contact. But violations can occur when proximity of units allows for contact, or when network-based interactions are present. This presents a particular concern for evaluation of engineering interventions, which are often assigned within networks in which improvements on one link or node can affect outcomes for other spatio-temporal units throughout the network. 

The paper has shown that the models available to infer causal effects have their own particular inferential assumptions, which must be rigorously evaluated prior to application. Crucially, the issue of identification must feature as a core considerations in formulating our conceptual and methodological approaches to causal problems in engineering research.

\bibliographystyle{chicago}
\bibliography{CausalReview}

\newpage
\section*{Appendix 1: Proofs }

\subsection*{Proof of the balancing property of the propensity score}

\begin{Proof}(Balancing property of the propensity score)\\  
First, in the case of a binary treatment 
\[\P(D_i=1|X_i,\pi(D_i=1|X_i))=\P(D_i=1|X_i)=\pi(D_i=1|X_i)\]
where the first equality follows from the fact that the PS is a function only of $X_i$ and the second from the definition of the PS.

Furthermore, by the law of iterated expectation we have that,
\begin{align*}
\mathbb{P}(D_i=1|\pi(D_i=1|X_i))&=\mathbb{E}[D_i|\pi(D_i=1|X_i)]=\mathbb{E}[\mathbb{E}[D_i|X_i,\pi(D_i=1|X_i)]|\pi(D_i=1|X_i)]\\
&=\mathbb{E}[\pi(D_i=1|X_i)|\pi(D_i=1|X_i)]=\pi(D_i=1|X_i)
\end{align*}
Hence we have that
\begin{align*}
\mathbb{P}(D_i=1|X_i,\pi(D_i=1|X_i))=\mathbb{P}(D_i=1|\pi(D_i=1|X_i))
\end{align*}
and thus
\begin{align*}
D_i \independent X_i |\pi(D_i=1|X_i)
\end{align*} 
For multivalued or continuous treatment we can apply the same logic. Let $\mathcal{X}$ be the sample space in which covariates $X_i$ lie, then 
\begin{align*}
f_{D|\pi}\left(d|\pi(d|x_i)\right)&=\int_{\mathcal{X}} f_{D,X|\pi}\left(d,x_i|\pi(d|x_i)\right)dx_i\\
&=\int_{\mathcal{X}} f_{D|X,\pi}\left(d|x_i,\pi(d|x_i)\right)f_{X|\pi}\left(x_i|\pi(d|x_i)\right) dx_i\\
&=\int_{\mathcal{X}} f_{D|X}(d|x_i)f_{X|\pi}\left(x_i|\pi(d|x_i) \right)dx_i\\
&=\int_{\mathcal{X}} \pi(d|x_i)f_{X|\pi}\left(x_i|\pi(d|x_i) \right)dx_i,\\
&= \pi(d|x_i)=f_{D|X}(d|x_i).
\end{align*}
Thus,$f_{D|\pi}\left(d|\pi(d|x_i)\right)=f_{D|X}(d|x_i)$, and therefore we have the property of balancing 
\begin{align*}
\Expect\left[I_{d}(D_i)|X_i, \pi(d|X_i)\right]=\Expect\left[I_{d}(D_i)|\pi(d|X_i)\right].
\end{align*} 
\end{Proof}

\subsection*{Proof of causal identification under strong ignorability via propensity score adjustment}

\begin{Proof}(Identification of an APO under strong ignorability via PS adjustment).  
\begin{subequations}
For binary treatments
\label{DRGPS}
\begin{align} 
\mu(1)= & \Expect\left[Y_i(1)\right]\\
= & \Expect_{X}\left[\Expect\left(Y_i(1)|X_i\right)\right]\\
= &\Expect_{X}\left[\Expect\left[Y_i(1)|\pi(D_i=1|X_i)\right]\right]\\
= &\Expect_{X}\left[\Expect\left[Y_i|D_i=1,\pi(D_i=1|X_i)\right]\right].
\end{align} 
\end{subequations}
For multivalued or continuous treatments, by conditional probability and corollary (\ref{Cor1}) above we have
\begin{align*}
f_{Y_{d,i}|D,\pi}(y_{d,i}|d,\pi(d|x_i))&= \frac{f_{Y_{d,i}|\pi}(y_{d,i}|\pi(d|x_i))f_{D|Y_{d,i},\pi}(d|Y_{d,i},\pi(d|x_i))}{f_{D|\pi}(d|\pi(d|x_i))} \\
&=\frac{f_{Y_{d,i}|\pi}(y_{d,i}|\pi(d|x_i))f_{D|\pi}(d|\pi(d|x_i))}{f_{D|\pi}(d|\pi(d|x_i))} \\
&=f_{Y_{d,i}|\pi}(y_{d,i}|\pi(d|x_i)),
\end{align*}
and therefore
\[\Expect[Y_i(d)|D=d,r(d,X_i)=\pi(d|x_i)]=\Expect[Y_i(d)|r(d,X_i)=\pi(d|x_i)].\]
However,
\begin{align*}
\Expect[Y_i(d)|D=d,\pi=\pi(d|x_i)]&= \Expect[Y_i(d)|D=d,r(D,X_i)=\pi(d|x_i)] \\
&= \Expect[Y_i(d)|D=d,r(d,X_i)=\pi(d|x_i)]\\
&= \Expect[Y_i(d)|r(d,X_i)=\pi(d|x_i)]\\
&=\beta(d,\pi(d|x_i)),
\end{align*}
and so by iterated expectations
\begin{equation} \label{DRGPS}
\mu(d)\equiv \Expect[Y_i(d)]\equiv \Expect_{X_i}[\Expect(Y_i(d)|X_i)]=\Expect_{X_{i}}[\Expect[Y_i(d)|\pi=\pi(d|x_i)]]=\Expect_{X_{i}}[\beta(d,\pi(d|x_i))].
\end{equation} 
\end{Proof} 

\subsection*{Proof of the Theorem of Doubly Robust Estimation}

The following proof of the DR property borrows from \citet{Naik/et/al:2016} and is a direct extension of that given by \citep{Lunceford/Davidian:2004} from the binary treatment case to the multivalued case. As such the proof in the binary case follows as a special case of this. 

\begin{Proof} (Double-robustness) 
For dose $d \in \mathcal{D} \{d_0,d_1,...,d_m\}$ we write the DR estimator for $\hat{\mu}_{DR}(d)$ as
\begin{align}\label{APmudr}
\hat{\mu}_{DR}(d)=\frac{1}{n}\sum_{i=1}^n\left[\Psi^{-1}\left\{m(d,X_i;\hat{\beta})\right\}+\frac{I_d(D_i)}{\hat{\pi}(d|X_i;\hat{\alpha})}\left[Y_i-\Psi^{-1}\left\{m(d,X_i;\hat{\beta})\right\}\right]\right].
\end{align}
Applying the WLLN to (\ref{APmudr})  we have
\begin{align} \label{DRwwln}
\hat{\mu}_{DR}(d)\overset{p}{\to} \Expect\left[Y\right]+ \Expect\left[\frac{I_d(D)-\pi(d|X;\alpha)}{\pi(d|X;\alpha)}\bigg\{Y-\Psi^{-1}\left\{m(d,X;\beta)\right\}\bigg\}\right],
\end{align}
which follows because 
\begin{align*}
\Expect\left[\frac{I_d(D)-\pi(d|X;\alpha)}{\pi(d|X;\alpha)} Y\right]=\Expect\left[\frac{I_d(D)Y}{\pi(d|X;\alpha)}\right]-\Expect\left[\frac{\pi(d|X;\alpha)Y}{\pi(d|X;\alpha)}\right],
\end{align*}
and where $\pi(d|X;\alpha)$ is the postulated PS model for $\pi(d|X)$,  $\Psi^{-1}\left\{m(d,X;\beta)\right\}$ is the postulated OR model, and $\alpha$ and $\beta$ are the ``true'' parameters of the PS and OR models respectively. 

From the SUTVA, $Y(d)=I_d(D)Y$, thus (\ref{DRwwln}) can be written
\begin{align} \label{DReq}
\Expect\left[Y(d)\right]+ \Expect\left[\frac{I_d(D)-\pi(d|X;\alpha)}{\pi(d|X;\alpha)}\bigg\{Y(d)-\Psi^{-1}\left\{m(d,X;\beta)\right\}\bigg\}\right],
\end{align}

The first term of (\ref{DReq}) is what we are trying to estimate, so all that remains is to show that the second term is 0 if either the PS or OR model is correctly specified. Let us first consider the case where the OR model is correctly specified, so that
\begin{align*}
\Psi^{-1}\left\{m(d,X;\beta)\right\}=\Expect(Y|D=d,X)
\end{align*}
Then, using the above equality and the law of iterated expectation, the second term of $(3.4)$ becomes
 \begin{align*}
&\Expect\bigg(\Expect\left[\frac{I_d(D)-\pi(d|X;\alpha)}{\pi(d|X;\alpha)}\bigg\{Y(d)-\Expect(Y|D=d,X)\bigg\}\bigg|I_d(D),X\right]\bigg)\\
=&\Expect\bigg(\frac{I_d(D)-\pi(d|X;\alpha)}{\pi(d|X;\alpha)}\Expect\left[\bigg\{Y(d)-\Expect(Y|D=d,X)\bigg\}\bigg|I_d(D),X\right]\bigg)\\
=&\Expect\bigg(\frac{I_d(D)-\pi(d|X;\alpha)}{\pi(d|X;\alpha)}\bigg\{\Expect[Y(d)|I_d(D),X]-\Expect(Y|D=d,X)\bigg\}\bigg)\\
=&\Expect\bigg(\frac{I_d(D)-\pi(d|X;\alpha)}{\pi(d|X;\alpha)}\bigg\{\Expect[Y(d)|X]-\Expect(Y(d)|X)\bigg\}\bigg)=0
\end{align*}
where the last line follows from the weak conditional independence condition implied by the no unmeasured confounders assumption, which gives that
\begin{align*}
\Expect(Y|D=d,X)=\Expect(Y(d)|D=d,X)=\Expect(Y(d)|X)=\Expect(Y(d)|I_d(D),X)
\end{align*}

Conversely, if the PS model is correctly specified then 
\begin{align*}
\pi(d|X;\alpha)=\pi(d|X)=\mathbb{P}(D=d|X)=\Expect(I_d(D)|X)
\end{align*}
Thus, using the above equality and the law of iterated expectation, the second term of $(3.4)$ becomes
\begin{align*}
&\Expect\bigg(\Expect\left[\frac{I_d(D)-\pi(d|X)}{\pi(d|X)}\bigg\{Y(d)-\Psi^{-1}\left\{m(d,X;\beta)\right\}\bigg\}\bigg|Y(d),X\right]\bigg)\\
=&\Expect\bigg(\bigg\{Y(d)-\Psi^{-1}\left\{m(d,X;\beta)\right\}\bigg\}\Expect\left[\frac{I_d(D)-\pi(d|X)}{\pi(d|X)}\bigg|Y(d),X\right]\bigg)\\
=&\Expect\bigg(\bigg\{Y(d)-\Psi^{-1}\left\{m(d,X;\beta)\right\}\bigg\}\frac{\Expect[I_d(D)|Y(d),X]-\pi(d|X)}{\pi(d|X)}\bigg)\\
=&\Expect\bigg(\bigg\{Y(d)-\Psi^{-1}\left\{m(d,X;\beta)\right\}\bigg\}\frac{\Expect[I_d(D)|X]-\pi(d|X)}{\pi(d|X)}\bigg)\\
=&\Expect\bigg(\bigg\{Y(d)-\Psi^{-1}\left\{m(d,X;\beta)\right\}\bigg\}\frac{\pi(d|X)-\pi(d|X)}{\pi(d|X)}\bigg)=0
\end{align*}
where the second last line follows from the no unmeasured confounders assumption as before.

Thus we see that $\hat{\mu}_{DR}(d)$ consistently estimates the APO $\mu(d)$ when either the OR model or PS model is correctly specified, in the case of multivalued treatment.
\end{Proof}

\subsection*{Identification of an APO using inverse Propensity score weighting}

\begin{Proof}(Identification of an APO using inverse Propensity score weighting)  
\begin{align*}\label{psw}
\Expect_i\left[ \frac{I_{d}(D_i) \cdot Y_i}{\pi(D_i|X_i;\alpha)}\right]&= \Expect_i\left[\frac{I_{d}(D_i) \cdot Y_i(1)}{\pi(D_i|X_i;\alpha)}\right]\\
&= \Expect_{X_i}\left[\Expect_i\left(\frac{I_{d}(D_i) \cdot Y_i(1)}{\pi(D_i|X_i;\alpha)}\biggr| X_i \right) \right]\\ 
&= \Expect_{X_i} \left[\frac{\Expect_i(I_{d}(D_i)|X_i)\cdot \Expect_i(Y_i(1)|X_i)}{\pi(D_i|X_i;\alpha)} \right]\\
&= \Expect_{X_i} \left[\frac{\pi(D_i|X_i;\alpha)\cdot \Expect_i(Y_i(1)|X_i)}{\pi(D_i|X_i;\alpha)} \right]\\ 
&=\Expect_{X_i}[\Expect_i(Y_i(1)|X_i)]\\
& =\Expect_i[Y_i(1)]=\mu(d)
\end{align*}
\end{Proof}

\section*{Appendix 2: Simulation figures}
\begin{figure}[htp]
\centering 
\includegraphics[height=6cm]{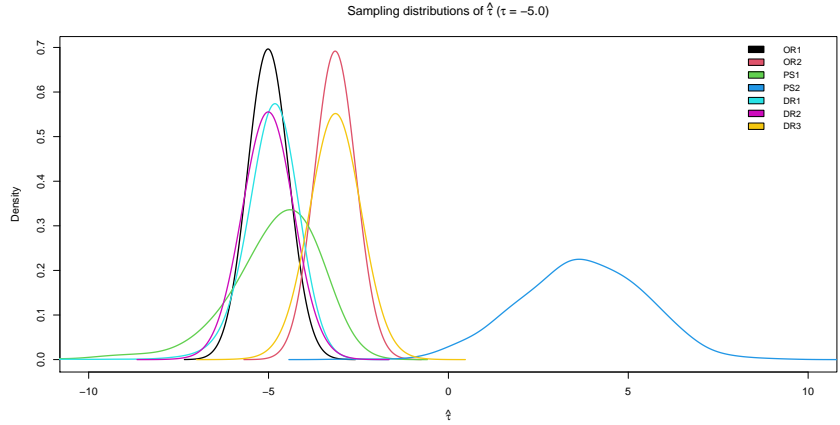}
{\caption{Simulation of causal ATE estimators under strong ignorability ($\tau=-5.0$)\label{CSSimF}}}
\end{figure}

\begin{figure}[htp]
\centering 
\includegraphics[height=6cm]{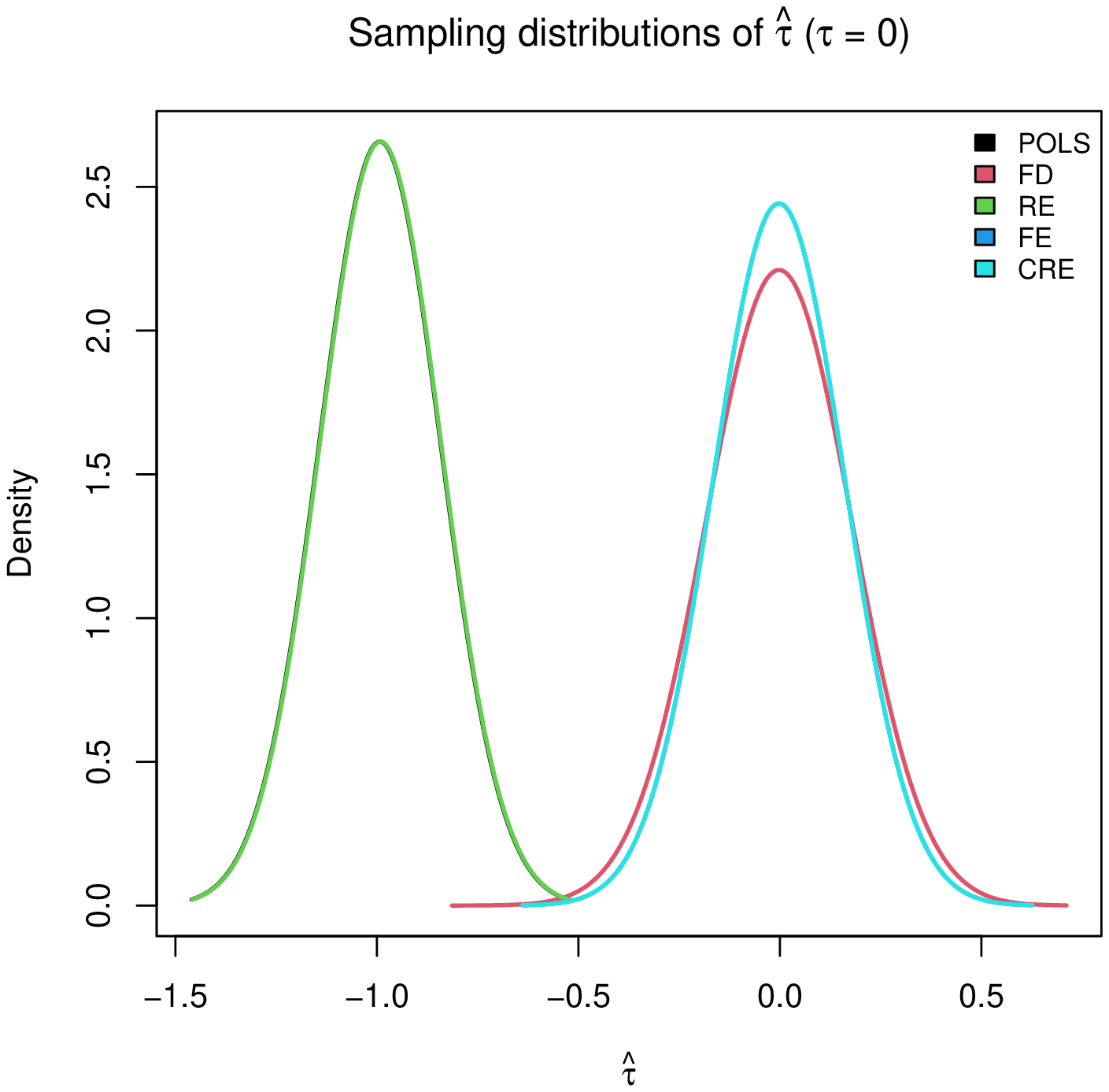}
{\caption{Simulation of panel ATE estimators with time invariant confounding ($\tau=2.0$)\label{PD1SimF}}}
\end{figure}

\begin{figure}[htp]
\centering 
\includegraphics[height=6cm]{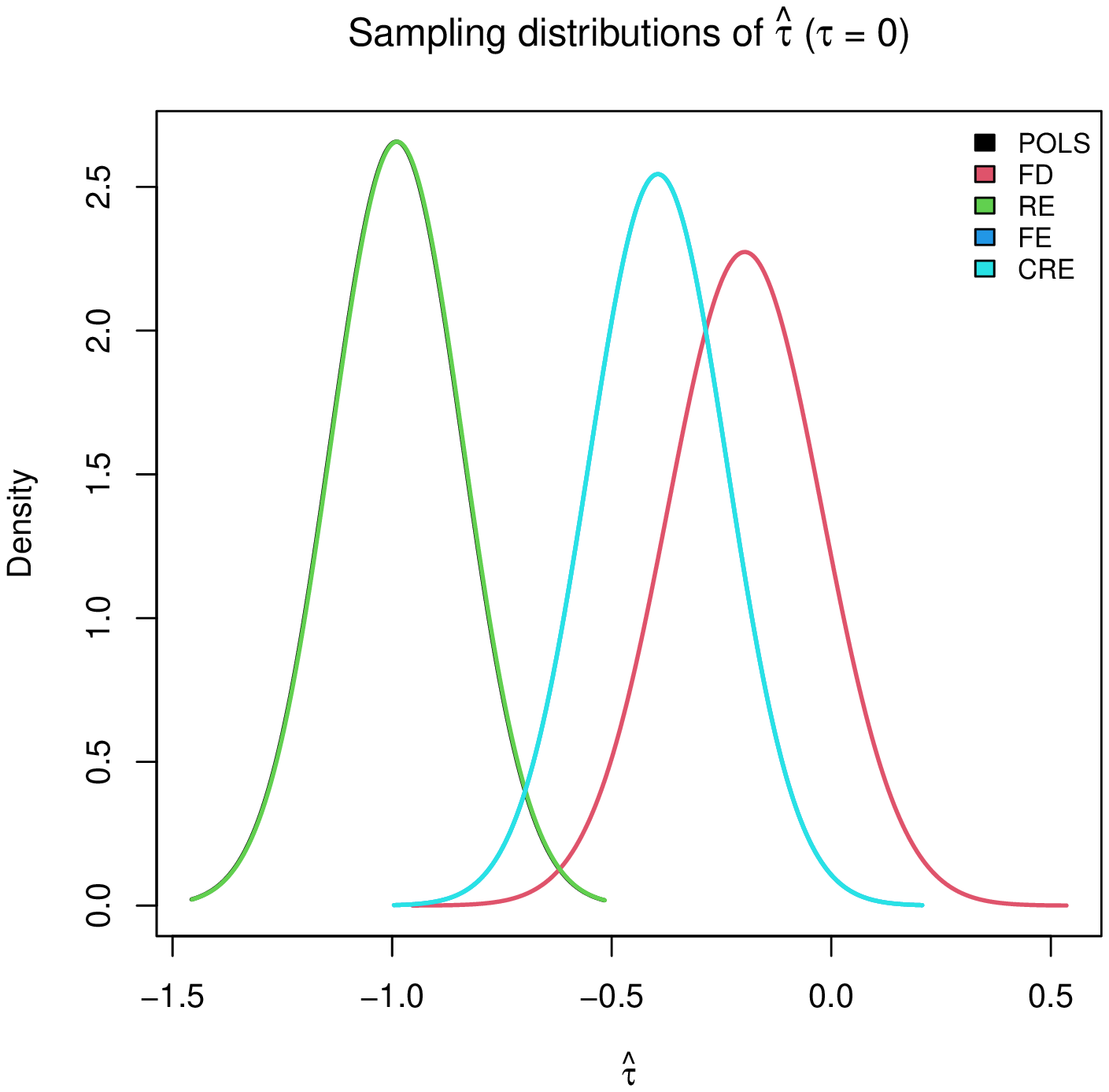}
{\caption{Simulation of panel ATE estimators with time varying confounding ($\tau=2.0$)\label{PD2SimF}}}
\end{figure}

\begin{figure}[htp]
\centering 
\includegraphics[height=6cm]{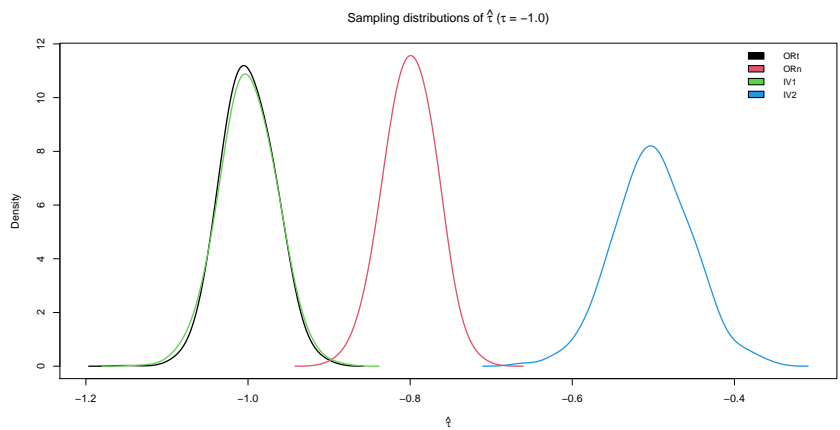}
{\caption{Simulation of IV estimator under non-ignorability ($\tau=1.0$)\label{IVSimF}}}
\end{figure}

\begin{figure}[htp]
\centering 
\includegraphics[height=6cm]{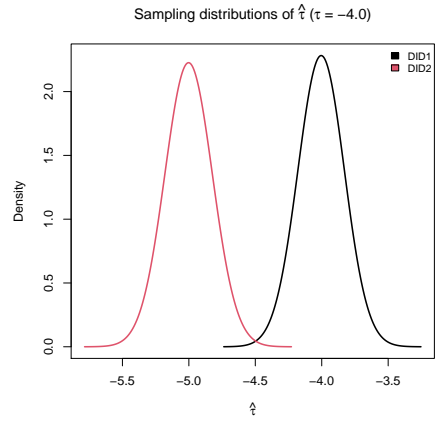}
{\caption{Simulation of DID estimator ($\tau=4.0$)
\label{DIDSimF}}}
\end{figure}

\begin{figure}[htp]
\centering 
\includegraphics[height=6cm]{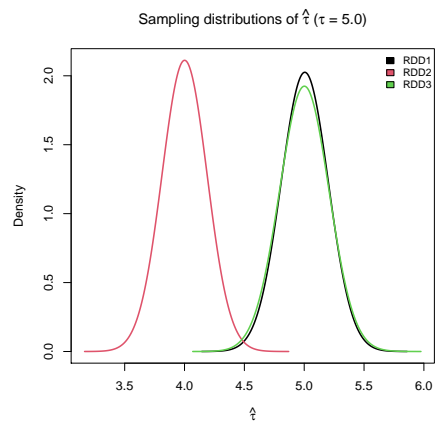}
{\caption{Simulation of RDD estimator ($\tau=5.0$)
\label{RDDSimF}}}
\end{figure}
\end{document}